\documentclass[onecolumn]{aastex631}
\usepackage{amsmath}
\usepackage{mathrsfs}
\usepackage{makecell}
\usepackage{multirow}
\received{...}
\revised{...}
\accepted{...}
\published{...}

\begin{document}

\title{Prospects of Identifying Hierarchical Triple Mergers for the Third-generation Ground-based Detectors}
\author{Bo Gao}
\affiliation{Key Laboratory of Dark Matter and Space Astronomy, Purple Mountain Observatory, Chinese Academy of Sciences, Nanjing 210033, China}
\affiliation{School of Astronomy and Space Science, University of Science and Technology of China, Hefei, Anhui 230026, China}
\author[0000-0001-9120-7733]{Shao-Peng Tang}
\affiliation{Key Laboratory of Dark Matter and Space Astronomy, Purple Mountain Observatory, Chinese Academy of Sciences, Nanjing 210033, China}
\correspondingauthor{Shao-Peng Tang}
\email{tangsp@pmo.ac.cn}
\author{Jingzhi Yan}
\correspondingauthor{Jingzhi Yan}
\email{jzyan@pmo.ac.cn}
\affiliation{Key Laboratory of Dark Matter and Space Astronomy, Purple Mountain Observatory, Chinese Academy of Sciences, Nanjing 210033, China}
\affiliation{School of Astronomy and Space Science, University of Science and Technology of China, Hefei, Anhui 230026, China}
\author[0000-0002-8966-6911]{Yi-Zhong Fan}
\affiliation{Key Laboratory of Dark Matter and Space Astronomy, Purple Mountain Observatory, Chinese Academy of Sciences, Nanjing 210033, China}
\affiliation{School of Astronomy and Space Science, University of Science and Technology of China, Hefei, Anhui 230026, China}

\begin{abstract}
A hierarchical triple merger (HTM) constitutes a type of event in which two successive black hole (BH) mergers occur sequentially within the observational window of gravitational wave (GW) detectors, which has important role in testing general relativity and studying BH population.
In this work, we conduct an analysis to determine the feasibility of identifying HTMs from a large GW event catalog using the third-generation ground-based GW detectors.
By comparing the Bhattacharyya coefficient that measures the overlap between the posterior distributions of the remnant and progenitor BH parameters, we find that the overlap between the event pair can serve as a preliminary filter, which balances between computational demand and the probability of false alarms.
Following this initial, time-efficient, yet less accurate screening, a subset of potential HTM candidates will be retained.
These candidates will subsequently be subjected to a more precise, albeit time-intensive, method of joint parameter estimation for verification.
Ultimately, this process will enable us to robustly identify HTMs. 
\end{abstract}

\keywords{Gravitational wave sources (667); Compact objects (288)}
\section{Introduction} \label{sec-intro}

The observation of gravitational waves (GWs) by the advanced LIGO/Virgo detectors (aLIGO/AdV) has opened new frontiers in the study of compact binary mergers, providing stringent tests of general relativity \citep{2021arXiv211206861T} and shedding light on the origins of coalescing binary black holes (BBHs) \citep{2023PhRvX..13a1048A}.
To date, aLIGO/AdV have cataloged over 90 GW events \citep{2021arXiv211103606T}, predominantly from BBH mergers.
Some of these BBHs, particularly those with masses within the predicted pair-instability supernova (PISN) mass gap, are postulated to originate from hierarchical merger processes \citep{2021NatAs...5..749G, 2022PhR...955....1M, 2023arXiv230302973L}.
This process may constitute a significant mechanism for the formation of intermediate-mass black holes in globular clusters \citep{2002MNRAS.330..232C}. Extensive research has been conducted on the hierarchical merger scenario, including studies on the modeling of BBH populations with hierarchical mergers \citep{2020ApJ...893...35D}, the distribution of spin magnitudes resulting from hierarchical mergers \citep{2017ApJ...840L..24F}, the coagulation processes within the dense environments of globular clusters that may explain mergers spanning the lower and upper mass gaps \citep{2021MNRAS.507..743F}, the differentiation between field and dynamical capture formation scenarios \citep{2017PhRvD..95l4046G}, and the simulation of hierarchical mergers across various environments \citep{2021Symm...13.1678M}.
The O4/O5 observation runs are expected to increase the inventory of compact binary coalescences, enhancing the detections of black holes that have formed through hierarchical mergers.
As the dataset of GW observations grows, we anticipate the possibility of detecting a new category of systems: hierarchical triple mergers (HTMs), i.e., two successive BH mergers occurred sequentially within the observation window of GW detectors \citep{2019MNRAS.482...30S, 2020MNRAS.498L..46V, 2021ApJ...907L..48V}.
Astrophysical sites such as star clusters and the accretion disks of active galactic nuclei (AGN) are considered conducive to such three-body mergers, given the higher probability of BBH encounters with solitary black holes (BHs) \citep{2019MNRAS.482...30S, 2022Natur.603..237S}.
Additionally, hierarchical triples are theorized to form in the dense cores of globular clusters via frequent binary-mediated interactions, which may lead to the creation of intermediate-mass black holes \citep{Liang:2017cjo,2023arXiv231105393L}.
The investigation of hierarchical triple mergers offers a wealth of scientific prospects.
They are crucial for testing the fundamental principles of general relativity \citep{2023MNRAS.523.4113T}, exploring spin-orbit dynamics \citep{2023arXiv230701903O}, probing the characteristics of the most massive neutron stars \citep{Tang:2023zxa} and then testing the results obtained in other approaches \citep{Fan:2023spm}, and providing alternative approaches to delineate the formation channels of coalescing compact binaries \citep{2022MNRAS.511.1362T}.

Therefore, it is essential to develop robust techniques for the identification of such systems.
The strategy for detecting HTMs parallels the approach used for identifying strong GW lensing signals, which involves determining the relationship between event pairs based on intrinsic and extrinsic parameters.
Although suspected strongly lensed event pairs have been reported, such as GW170104 and GW170814 \citep{2021ApJ...908...97L}, no confident gravitational lensing of GW events was confirmed during the three observing runs of aLIGO/AdV \cite{2021ApJ...923...14A,2023arXiv230408393T}.
The European Einstein Telescope (ET) \citep{2010CQGra..27s4002P, 2011CQGra..28i4013H} and the American Cosmic Explorer (CE) \citep{2019BAAS...51g..35R, 2021arXiv210909882E} promise an order of magnitude improvement in sensitivity and enhanced bandwidth for the upcoming third-generation (3G) era. 
This advancement in detection capabilities will likely reduce the uncertainty in parameter measurement and facilitate the identification of novel events, including strongly lensed GW signals and HTMs.
For the search of event pairs, a relatively rapid method is to assess posterior overlap by comparing the posteriors derived from individual analyses.
However, this method carries the risk of false positives when overlaps occur by chance \citep{2018arXiv180707062H, 2023MNRAS.519.2046J}.
An alternative, more precise technique is joint parameter estimation (JPE), which assumes a relationship between parameters and analyzes two signals concurrently.
Despite its higher accuracy, JPE is time-consuming and has been employed in analyses involving overlapping GW and strongly lensed GW signals \citep{2023MNRAS.526.3088J, 2023MNRAS.523.1699J}.
As the number of detected events grows, the rapid increase in event pairs poses significant computational challenges for the JPE method and increases the probability of false alarms for the overlap method.
A combined approach could mitigate these issues \citep{2023arXiv230408393T}. 
A high rate of false positives in the overlap posterior would necessitate excessive follow-up JPE analyses.
This problem is expected to intensify in the 3G detector era, as the number of GW events is projected to reach the order of $10^5$ and the detected GW signals will have a longer duration.
Consequently, there is a pressing need for an improved overlap method for preliminary screening.

In this study, we propose a two-step approach to identify HTM signatures using 3G GW detectors.
The first stage involves assessing the similarity between posterior distributions of mock event pairs generated from a specified BBH population model.
This preliminary examination is designed to filter out candidate event pairs with a predefined criterion of similarity, specifically concerning the characteristics of BH remnants in the first merger compared to the BH properties in the second merger.
Subsequent verification of genuine HTM signals is achieved through a JPE for each candidate event pair, conducted within a Bayesian parameter inference framework.
Through model comparison, it is robust to evaluate whether an event pair is related to HTM.
Our approach then will facilitate the accurate discernment of HTMs within a substantial dataset encompassing ten thousand events.

\section{Method}
\subsection{Preparing mock event data}

\begin{figure}
    \centering
    \includegraphics[width=0.98\textwidth]{"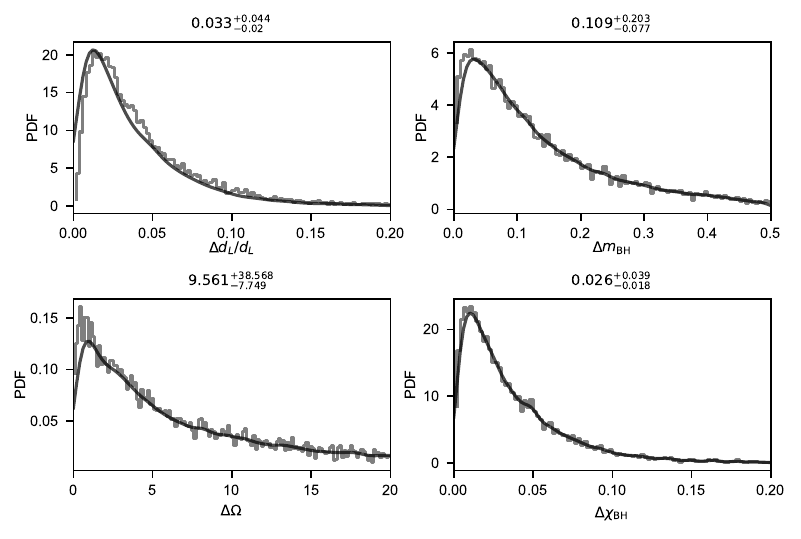"}
    \caption{Distributions of measurement uncertainties for the luminosity distance $d_L$, the sky location $\Delta \Omega$ (The units is $\rm deg ^2$ with $90\%$ probability), and the mass $m_{\rm BH}$ and spin magnitude $\chi_{\rm BH}$ of the primary black hole. The black solid lines represent the Kernel Density Estimates (KDEs) of the distributions. The values reported in each plane are the median and $68.3\%$ symmetric credible interval.}
    \label{fig:pop_un}
\end{figure}

We simulate BBH mergers using a population model informed by data from the LIGO/Virgo collaboration (as outlined in Table~\ref{tab:popu}). In Fig.~\ref{fig:pop_un}, we showcase the distributions (with signal-to-noise $\geq 12$) of measurement uncertainties for primary black hole's mass and spin.
The HTM scenario is constructed as follows:
First, a merger is randomly selected from the simulated sample to represent the initial merger in the HTM.
The outcome of this first merger establishes the characteristics of the remnant BH, which then serves as one component in the subsequent merger.
Thereafter, we sample the properties of a companion BH from the same population model to constitute the second BBH system.
The second merger is assumed to take place one year after the initial merger.
We have generated two distinct event catalogs characterized by differing proportions of HTM events.
The first, termed the `Test Catalog', is designed exclusively to assess the efficacy of four screening methods (detailed in next subsection).
It is important to note that this catalog may not accurately represent the true astrophysical distribution of HTMs.
The second, known as the `Fiducial Catalog', is constructed based on contemporary understanding of the BBH population.
According to \citet{2023MNRAS.523.4113T}, HTM events are estimated to comprise approximately 0.1\% of detectable BBHs.
Consequently, the `Fiducial Catalog' has an ensemble of around ten thousand BBH mergers, which includes ten HTM events.
We generate a sample of $z$ according to the redshift distribution and subsequently convert it to luminosity distance by assuming a flat $\Lambda\mathrm{CDM}$ cosmology based on PLANCK18 parameters \citep{2020A&A...641A...6P}.
The anticipated number of detectable BBH mergers is projected to be approximately $\mathcal{O}(10^4)$ during the era of 3G GW detectors.

The mock event catalog is extracted from the simulated mergers by excluding events where the optimal signal-to-noise ratio (S/N) falls below the threshold of 12.
With this catalog in hand, we proceed to assign mock measurement uncertainties to the parameters of the events.
To estimate these uncertainties, we employ the novel Fisher-matrix code GWFAST \citep{Iacovelli:2022mbg}.
Our analysis is conducted within the context of a network comprising three 3G ground-based detectors, specifically the ET and two CE detectors.
The configuration of these detectors aligns with the settings adopted in \citet{Iacovelli:2022bbs}.
For the calculation of the Fisher Information Matrix (FIM) and the S/N, we utilize the waveform model IMRPhenomXPHM \citep{2021PhRvD.103j4056P}.
Events with an inversion error of the FIM exceeding 0.05 are also discarded, as they do not meet the reliability criteria for parameter error estimation.
Having determined the measurement uncertainties, we then assign mock median parameter values for each `observation'.
These values are drawn from a joint normal distribution with the mean corresponding to the injected parameters and the covariance corresponding to the inverse of FIM, which accounts for the correlation between mass and spin parameters.
This process is referred to as the `shift value' strategy in this work. 
When it comes to the mock observations of the remnant BHs, we note that their parameter distribution can be derived from the properties of the progenitor BHs using the functions --- final\_mass\_from\_initial and final\_spin\_from\_initial in PYCBC \citep{2019PASP..131b4503B}.
Thus, the mock median values and uncertainties for the remnant mass and spin can be directly obtained with the parameter distributions.

\begin{deluxetable*}{ccc}\label{tab:popu}
    \tablecaption{Distributions Used to Generate Mock BBH Events}
    \tablewidth{0pt}
    \tablehead{\colhead{Parameters} & \colhead{Description} & \colhead{Distributions} }
    \startdata
    $m_1$ & primary mass & \multirow{2}{*}{POWER LAW + PEAK\citep{2019ApJ...882L..24A}} \\
    $m_2$ & secondary mass & \\
    $z$ & redshift & Madau-Dickinson \\
    $d_L$ & luminosity distance & PLANCK18 flat $\Lambda$CDM \\
    $\chi_{1,2}$ & spin magnitude of object 1,2 & Beta distribution \citep{2019ApJ...882L..24A} \\
    $\theta_{1,2}$ & spin tilt of object 1,2 & Mixture of isotropic and aligned component \citep{2019ApJ...882L..24A}\\
    $\phi_{\rm JL}$ & \makecell{azimuthal angle between orbital \\ and total angular momentum} & Uniform in $[0, 2\pi]$ \\
    $\phi_{1,2}$ & \makecell{difference in azimuthal angle \\ between the spin vectors} & Uniform in $[0, 2\pi]$ \\
    $\cos{\theta}$ & sky position $\theta=\pi/2-\delta$ & Uniform in $[-1, 1]$ \\
    $\phi$ & sky position $\phi=\alpha$ & Uniform in $[0, 2\pi]$ \\
    $\theta_{\rm JN}$ & \makecell{inclination angle w.r.t \\ total angular momentum} & Uniform in $[0, \pi]$ \\
    $\psi$ & polarization angle & Uniform in $[0, \pi]$ \\
    $t_c$ & time of coalescence & Uniform in 10 yr \\
    $\Phi_c$ & phase at coalescence & Uniform in $[0, 2\pi]$
    \enddata
\end{deluxetable*}

\subsection{Methods to screen HTM events}\label{subsec:overlap}

For a given pair of HTM events, the mergers are characterized by identical sky locations and luminosity distances.
Furthermore, the mass and spin characteristics of one of the BH in the second merger coincide with those of the BH remnant from the first merger.
In other words, the posterior distributions of the relevant parameters for the two events are expected to exhibit a high degree of similarity.
As such, the similarity in posterior distributions serves as a criterion for distinguishing between HTM and non-HTM events.
Nonetheless, a high degree of similarity may occur by chance in non-HTM events, leading to mis-identification.
Consequently, the identification process necessitates a more precise methodology (detailed in the next subsection).
Various methods for calculating this overlap are available in the literature \citep{2012PatRe..45.1386L, Bazn2019QuantitativeAO, article2D}, each with differing efficacy in differentiating between the two types of event pairs under consideration.

In this study, we evaluate four methods designed to quantify the overlap between posterior distributions.
These methods include: the cumulative distribution function (CDF) method ($\mathscr{M}_{\rm CDF}$), the intersection area method ($\mathscr{M}_{\rm Intersection}$), the integral of the possibility distribution function method ($\mathscr{M}_{\rm Integral}$), and the Bhattacharyya coefficient method ($\mathscr{M}_{\rm Bhattacharyya}$). 
The $\mathscr{M}{\rm CDF}$ method calculates the dissimilarity between two distributions by taking the maximum absolute value of the difference in their CDFs.
Conversely, we define similarity as the complement of dissimilarity, approaching one for identical distributions.
The $\mathscr{M}_{\rm Intersection}$ method quantifies similarity based on the area of intersection between two distribution curves.
The $\mathscr{M}_{\rm Integral}$ method define overlap as \citep{2018arXiv180707062H, 2018ApJ...860....6A}
\begin{eqnarray}
    \mathcal{I}_\theta \equiv \int \frac{p(\theta|d_1)p(\theta|d_2)}{p(\theta)} \mathrm{d}\theta,
\end{eqnarray}
where $p(\theta)$ represents the prior distribution for the parameter $\theta$, whereas $p(\theta|d_{1, 2})$ signifies the posterior distribution resulting from either event 1 or event 2.
Here, we assume a uniform prior $p(\theta)$ when assessing the degree of overlap.
\begin{figure}
    \centering
    \includegraphics[width=0.98\textwidth]{"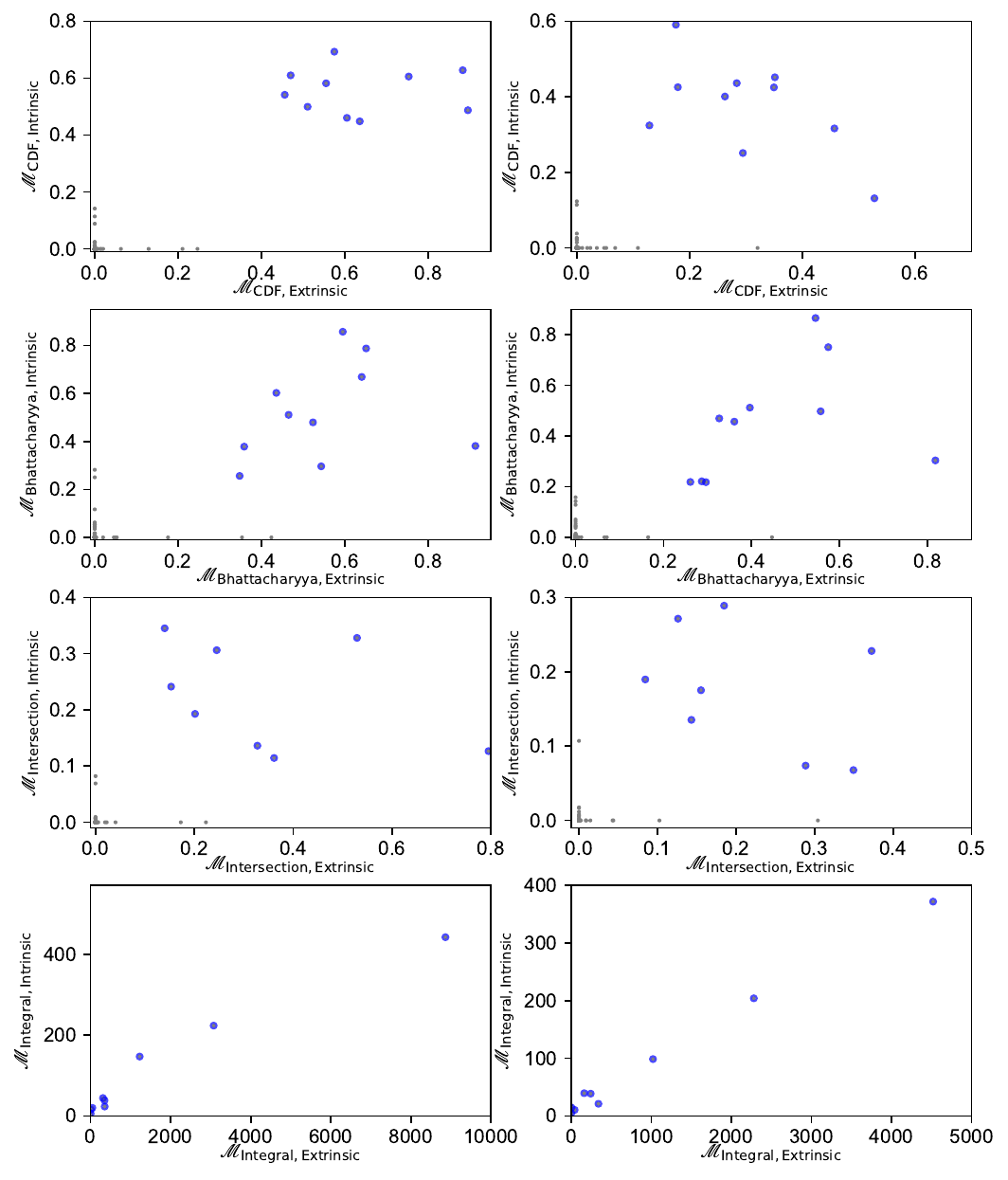"}
    \caption{
    Overlaps of intrinsic (vertical axis) and extrinsic (horizontal axis) parameters between two events calculated with four methods.
    For $\mathscr{M}_{\rm Bhattacharyya}$ method, we can directly derive the overall overlap for all extrinsic (intrinsic) parameters.
    In contrast, the methods $\mathscr{M}_{\rm CDF}, \mathscr{M}_{\rm Intersection}$, and $\mathscr{M}_{\rm Integral}$ involve calculating the overlap with respect to a single parameter.
    Here, the extrinsic overlap is modeled as the product of three extrinsic parameters, i.e., luminosity distance and sky location, whereas the intrinsic overlap is the product of two intrinsic parameters, namely mass and spin.
    The gray points denote all event pairs, while blue hollow circles distinguish the mixed HTMs.
    The left panels present scenarios without a mean value shift, whereas the right panel incorporates a mean value shift.
    The bottom row focuses on the overlap for the $\mathscr{M}_{\rm Integral}$ method, which exhibits an unbounded range of values.
    }
    \label{fig:overlap-method}
\end{figure}
The Bhattacharyya coefficient is a well-established metric for quantifying the degree of similarity between two probability distributions \citep{bhattacharyya1946measure}.
To calculate the Bhattacharyya distance between two probability distributions $P(x)$ and $Q(x)$, defined over a common domain $\pi$, we use the following expression,
\begin{eqnarray}
    {\rm D_B}(P, Q) = -\ln{{\rm BC}(P, Q)},
\end{eqnarray}
where 
\begin{eqnarray}
    {\rm BC}(P, Q) = \Sigma_{x\in\pi} \sqrt{P(x)Q(x)}.
\end{eqnarray}
For two multivariate normal distributions $\mathcal{N}(\boldsymbol{\mu}_i, \boldsymbol{\Sigma}_i)$, the Bhattacharyya distance can be written as
\begin{eqnarray}
    {\rm D_B}(p_1, p_2) = \frac{1}{8}\left(\bf{\mu_1} - \bf{\mu_2} \right)^{T} \Sigma^{-1} \left( \boldsymbol{\mu_1} - \boldsymbol{\mu_2} \right) + \frac{1}{2} \ln{\left( \frac{\mbox{det} \Sigma}{\sqrt{\mbox{det}\Sigma_1 \mbox{det}\Sigma_2}} \right)},
\end{eqnarray}
where $\Sigma = \frac{\Sigma_1+\Sigma_2}{2}$. And the Bhattacharyya coefficient, defined as,
\begin{eqnarray}
    \rho(P, Q) = e^{-{\rm D_B}(P, Q)},
\end{eqnarray}
serves a similar purpose in quantifying the extent of overlap between two samples or distributions.
The degree of overlap, as determined by $\mathscr{M}_{\rm CDF}, \mathscr{M}_{\rm Intersection}$, and $\mathscr{M}_{\rm Bhattacharyya}$ is confined within the range of $[0, 1]$.
Conversely, the overlap measured by $\mathscr{M}_{\rm Integral}$ is not limited, possessing an unbounded range.
A greater extent of overlap correlates with a higher degree of similarity between the two distributions.
An overlap value with a fixed range is more advantageous for establishing a criterion to successfully identify HTM events.

In this context, we categorize the mass and spin of a BH as intrinsic parameters, while the sky location and luminosity distance are considered as extrinsic parameters.
Among the methods evaluated, the overlap measured by the Bhattacharyya coefficient exhibits superior performance in distinguishing between HTM and non-HTM events (see the Results Section~\ref{sec:results}).
Consequently, this method is adopted for the initial screening process, which involves the following steps: Firstly, each event pair is ordered based on their coalescence times. Subsequently, the overlap of extrinsic parameters is computed, and the overlaps of mass and spin parameters are evaluated between the remnant of the first merger and each BH in the second merger, with the higher of the two overlaps being selected.

\subsection{Joint Bayesian analyses}\label{subsec:JPE}

The methods introduced in Section~\ref{subsec:overlap} exhibit expedience in screening numerous merger events; however, it incurs a high rate of misclassification.
A more reliable approach is necessary to validate that the selected events are indeed true HTM candidates.
Joint Parameter Estimation (JPE) analysis, grounded in Bayesian inference, offers greater precision in analyzing multiple events, albeit with a substantial computational demand.
Direct application of the JPE method to all event pairs would exceed our computational capabilities.
Therefore, we have derived a candidate set with manageable size, for which the computational demands of applying JPE remain within our reach.
In our Bayesian analyses, we employ the IMRPhenomXPHM waveform model, incorporating both effects of the spin-precession and higher harmonics.
In the era of 3G GW detectors, the extended signal duration, coupled with enhanced detector sensitivity, imposes a heavier computational load on Bayesian analysis.
To expedite parameter estimation, we have adopted the relative binning method proposed by \citet{2018arXiv180608792Z}, which has been integrated into the Bilby software \citep{2023arXiv231206009K}.
Additionally, we utilize the same approach as provided by PYCBC to discern the relationship between the progenitor and the remnant of the merger.
The Bayesian evidence for HTM hypothesis $\mathcal{H}_H$ can be calculated by
\begin{eqnarray}
    p(d_1, d_2|\mathcal{H}_H) = \int p(d_1|\boldsymbol{\theta_1}, \boldsymbol{\theta_{\rm ext}})p(d_2|\boldsymbol{\theta_2}, \boldsymbol{\theta_{\rm ext}}, m_f, a_f)p(\boldsymbol{\theta_1}, \boldsymbol{\theta_2}, \boldsymbol{\theta_{\rm ext}}) \mathrm{d}\boldsymbol{\theta_1}\mathrm{d}\boldsymbol{\theta_2}\mathrm{d}\boldsymbol{\theta_{\rm ext}},
\end{eqnarray}
where $\boldsymbol{\theta_{1,2}}$ represents GW parameters for individual event and $\boldsymbol{\theta_{\rm ext}}$ represents shared parameters, i.e., luminosity distance and sky location. The mass and spin for one of the component star in the second merger is set to $m_f$ and $a_f$, which are the remnant mass and spin of the first merger predicted by the waveform model. The Bayesian evidence for non-HTM hypothesis $\mathcal{H}_N$ is
\begin{eqnarray}
    p(d_1, d_2| \mathcal{H}_N) &=& \int p(d_1|\boldsymbol{\theta_1})p(d_2|\boldsymbol{\theta_2})p(\boldsymbol{\theta_1})p(\boldsymbol{\theta_2})\mathrm{d}\boldsymbol{\theta_1}\mathrm{d}{\boldsymbol{\theta_2}} \\
    &=& \int p(d_1|\boldsymbol{\theta_1})p(\boldsymbol{\theta_1})\mathrm{d}\boldsymbol{\theta_1} \times \int p(d_2|\boldsymbol{\theta_2})p(\boldsymbol{\theta_2})\mathrm{d}\boldsymbol{\theta_2} .
\end{eqnarray}
The Odds ratio between the two hypotheses $\mathcal{H}_H, \mathcal{H}_N$ is
\begin{eqnarray}
    \mathcal{O}_{\mathcal{H}_N}^{\mathcal{H}_H} = \frac{p(\mathcal{H_H}|d_1, d_2)}{p(\mathcal{H}_N|d_1, d_2)} = \frac{p(d_1, d_2|\mathcal{H}_H)}{p(d_1, d_2|\mathcal{H}_N)}~ \frac{p(\mathcal{H}_H)}{p(\mathcal{H}_N)}.
    \label{eq:odds}
\end{eqnarray}
The first ratio $p(d_1, d_2|\mathcal{H}_H)/p(d_1, d_2|\mathcal{H}_N)$, commonly referred to as the Bayes factor, can be obtained from Bayesian inference.
The second ratio $p(\mathcal{H}_H)/p(\mathcal{H}_N)$ is the prior odds of the two hypotheses.
We have assigned this prior odds a value of $0.1\%$, which corresponds to the proportion of HTM events within the total population of detectable mergers in the optimistic scenario \citep{2023MNRAS.523.4113T}.
Typically, an Odds ratio exceeding $10$ suggests that the data provide stronger support for one hypothesis over the other.

\section{Results}\label{sec:results}
\begin{figure}
    \centering
    \includegraphics[width=0.98\textwidth]{"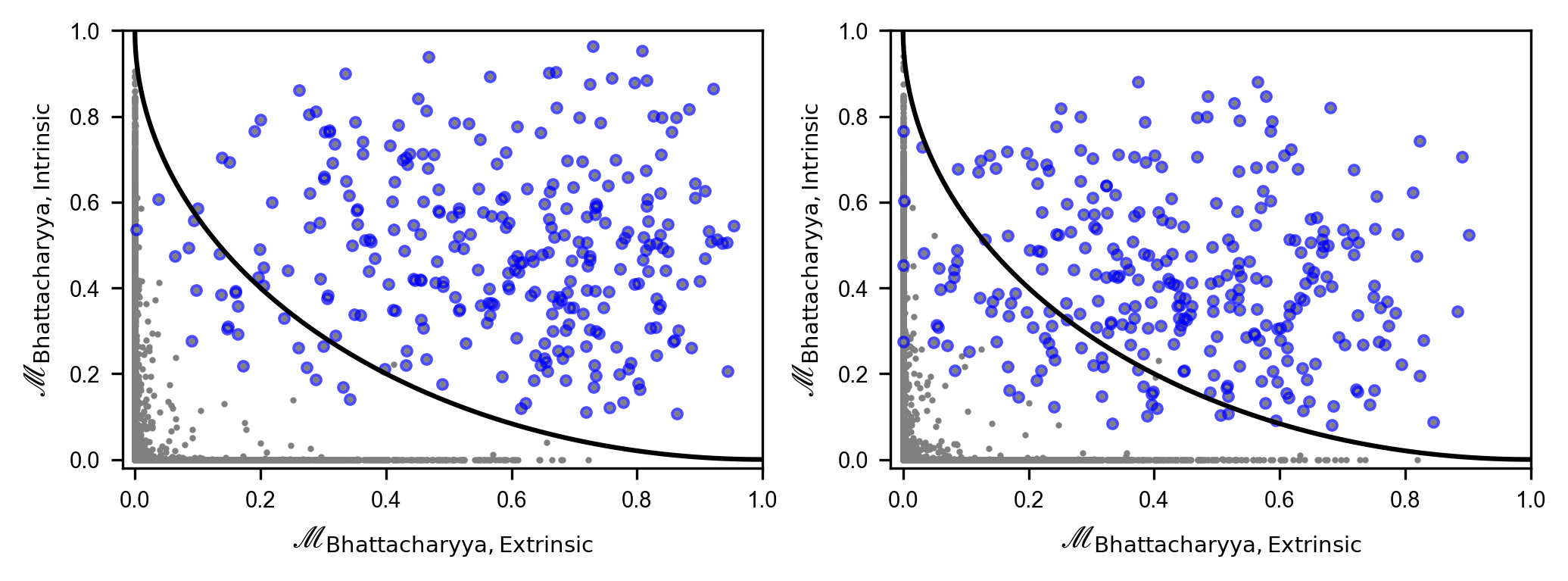"}
    \caption{Overlaps of intrinsic (vertical axis) and extrinsic (horizontal axis) parameters calculated with the Bhattacharyya coefficient method for the `Test Catalog(B)' that includes about 3000 events.
    The gray points denote all event pairs, while blue hollow circles distinguish the mixed HTMs.
    The left panels present scenarios without a mean value shift, whereas the right panel incorporates a mean value shift.
    A quarter of circle with a radius of 1.0 at $(1.0, 1.0)$, which serves as the overlap criterion, delineates the boundary distinguishing between HTM and non-HTM events.}
    \label{fig:identify}
\end{figure}
\begin{figure}
    \centering
    \includegraphics[width=0.98\textwidth]{"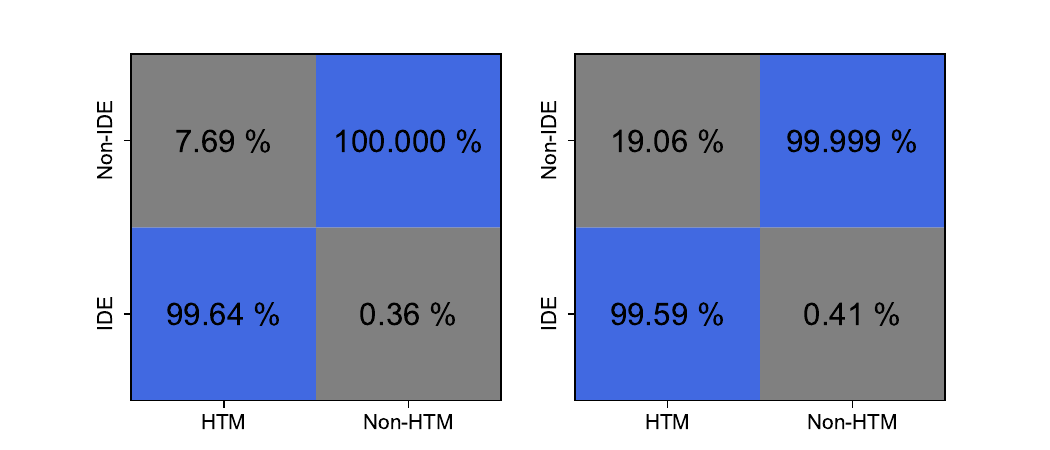"}
    \caption{Classification rates for the `Test Catalog(B)' that includes about 3000 events.
    The blue segment represents true positive rate (lower left part) and true negatives rate (top right part), whereas the gray segment denotes false positive rate (lower right part) and false negative rate (top left part).
    The designations `HTM' and `Non-HTM' refer to HTM events and non-HTM events, respectively. Similarly, `IDE' and `Non-IDE' distinguish between divisions into HTM and non-HTM events. The left panels present scenarios without a mean value shift, whereas the right panel incorporates a mean value shift.}
    \label{fig:rate}
\end{figure}
\begin{figure}
    \centering
    \includegraphics[width=0.98\textwidth]{"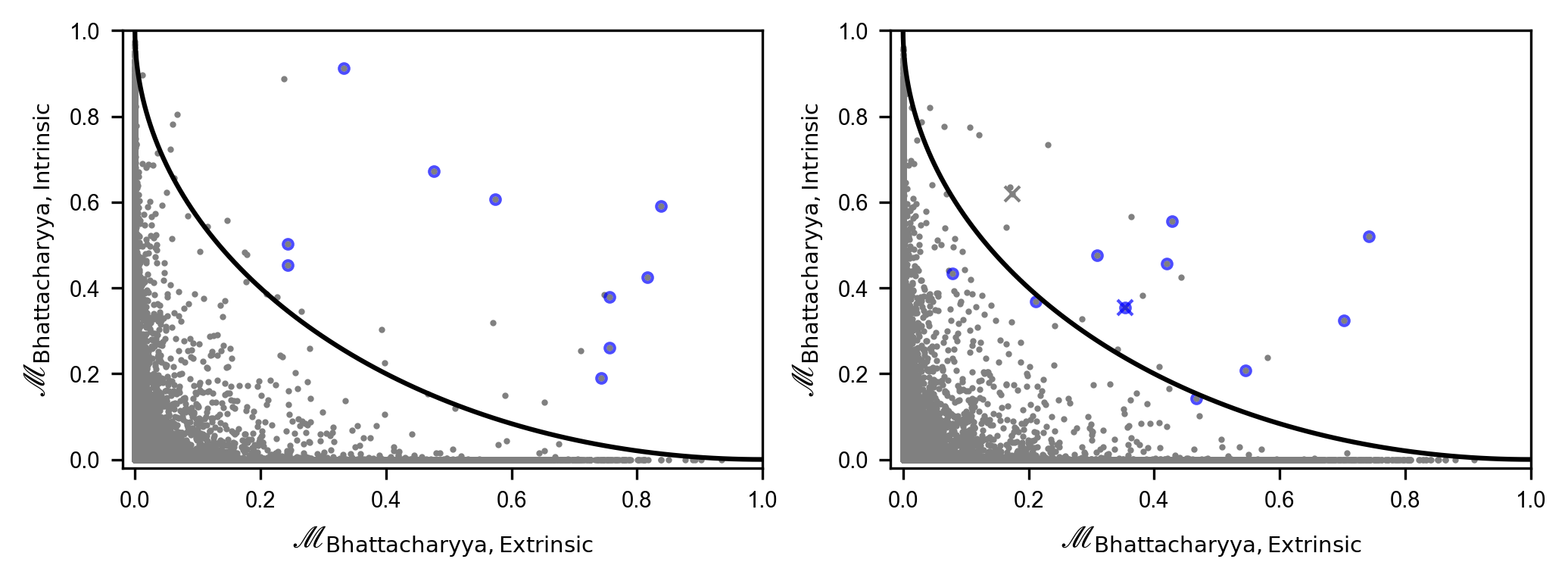"}
    \includegraphics[width=0.98\textwidth]{"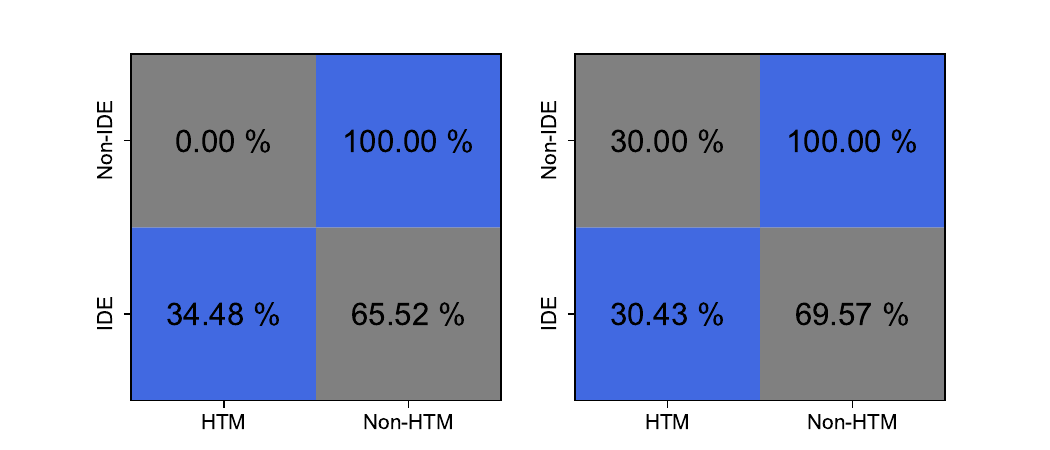"}
    \caption{Similar to Fig.~\ref{fig:identify} and Fig.~\ref{fig:rate}, but is for the `Fiducial Catalog'. The symbol $\times$ in the right panel denotes events selected for the JPE analyses.}
    \label{fig:overlap-mock}
\end{figure}

We first examine the performance of the four methods (introduced in Section.~\ref{subsec:overlap}) for calculating overlap within a `Test Catalog(A)' that has a HTM to non-HTM ratio of $1:10$ and a sample size of $100$.
As shown in Fig.~\ref{fig:overlap-method}, we find that the majority of non-HTM data points (gray) cluster near the coordinate axes, while HTM data points (blue) are predominantly located in the top right corner.
An analysis of the figure's left and right columns reveals that the overlap of extrinsic and intrinsic parameters more effectively distinguishes HTM when using the best-fit values without deviation.
Our findings indicate that all four methods can differentiate between HTM and non-HTM to varying degrees.
However, the $\mathscr{M}_{\rm Integral}$ method lacks a fixed range for overlap, which complicates the establishment of criteria for HTM identification.
And the $\mathscr{M}_{\rm Intersection}$ method exhibits a higher degree of blue and gray point intermixing, indicating poorer discrimination.
Upon comparing the $\mathscr{M}_{\rm CDF}$ and $\mathscr{M}_{\rm Bhattacharyya}$ methods, the latter provides a clearer separation between HTM and non-HTM, suggesting a superior capability for HTM identification.
Consequently, we have selected the Bhattacharyya method for preliminary screening.

To effectively pre-screen the mock catalog, it is essential to ascertain the overlap criteria that classifies HTM and non-HTM.
We apply the Bhattacharyya method to a `Test Catalog(B)' that has a HTM to non-HTM ratio of $1:10$ and a larger sample size of about $3000$.
The primary purpose of this `Test Catalog(B)' is to establish a demarcation between HTM and non-HTM based on the overlap of extrinsic and intrinsic parameters.
As illustrated in Fig.~\ref{fig:identify}, HTM data points gravitate away from the coordinate axes towards the upper right quadrant in the $\mathscr{M}_{\rm Bhattacharyya, Extrinsic}$ versus $\mathscr{M}_{\rm Bhattacharyya, Intrinsic}$ plane, while non-HTM points remain close to the axes, approaching $(0.0, 0.0)$ due to their negligible overlap.
The clear division between HTM and non-HTM by the $\mathscr{M}_{Bhattacharyya}$ method confirms the efficacy of the pre-screening process.
Based on the distribution within the $\mathscr{M}_{\rm Bhattacharyya, Extrinsic}$ versus $\mathscr{M}_{\rm Bhattacharyya, Intrinsic}$ plane, we can establish a criterion to categorize HTM.
For instance, event pairs located at a distance less than $1.0$ from the point $(1.0, 1.0)$ are classified as HTM candidates, as delineated by the black line in Fig.~\ref{fig:identify}.
The classification effectiveness is detailed in Fig.~\ref{fig:rate} according to the specified criteria.
The pre-screening process yields a true positive rate of $99.59\%$ and a false positive rate of $0.41\%$ among the HTM candidates.
The false negative rate among all mixed HTM is $19.06\%$.
Thus, the pre-screening demonstrates a robust capability to discern HTM in the `Test Catalog(B)', missing only $19.06\%$ of HTM at a low false alarm rate, while successfully filtering out approximately $99.999\%$ of non-HTM.
The analysis further reveals that utilizing best-fit values without deviation can reduce both the false alarm and false negative rates for HTM classification.

\begin{figure}
    \centering
    \includegraphics[width=0.68\textwidth]{"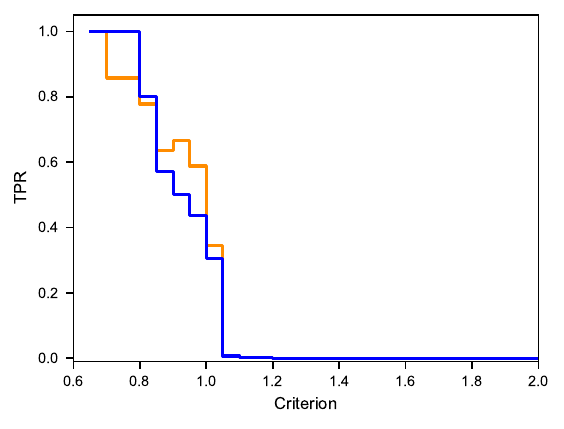"}
    \caption{True positive rate (TPR) of candidates as a function of the selection criterion applied during the initial screening phase for the `Fiducial Catalog'. This criterion is defined as the radius of a circle centered at the point (1.0, 1.0) in Fig.~\ref{fig:identify}. The orange and blue lines represent the results with and without deviation in the best-fit values, respectively. Notably, the TPR approaches zero when the threshold criterion is excessively large, because of too much inclusion of non-HTM.}
    \label{fig:threh}
\end{figure}

With the pre-screening criteria in hand, we apply the Bhattacharyya method to a `Fiducial Catalog' that has a HTM to non-HTM ratio of $1:1000$ (assuming an optimistic HTM detection rate) and a sample size of about $10000$ (about $\mathcal{O}(10^7)$ event pairs).
This sample size approximates the annual observational count anticipated in the 3G GW detector era.
The efficacy of HTM discrimination within the `Fiducial Catalog' is illustrated in Fig.~\ref{fig:overlap-mock}.
Following the initial screening, we identify a subset of 23 candidates, of which 7 are HTMs, from the initial $\mathcal{O}(10^7)$ event pairs.
Within this subset, the screening process achieved a true positive rate of $30.43\%$ and a false positive rate of $69.57\%$.
When comparing to the application within the `Test Catalog(B)', there is a significant decrease in the true positive rate.
This is because the `Fiducial Catalog', with its up to $\mathcal{O}(10^7)$ event pairs, may inadvertently include a higher proportion of remained non-HTMs in the candidates compared to the `Test Catalog(B)'.
The false negative rate is $30.00\%$, indicating that the pre-screening process successfully identify $70.00\%$ of all HTMs.
In summary, a threshold criterion of 1.0 demonstrates moderate effectiveness in identifying HTMs.
Varying the criterion, such as adjusting the radius of the circle used in the screening process, can yield a different size of candidate sets.
A higher threshold is advantageous for obtaining a smaller, more precise set of candidates with fewer non-HTMs, whereas a lower threshold is more inclusive, potentially capturing a greater number of HTMs.
The variation in the true positive rate with the threshold value is depicted in Fig.~\ref{fig:threh}.
As indicated, the true positive rate exceeds $80\%$ when the threshold value is set below 0.8 for the `Fiducial Catalog'.
Nevertheless, the possibility of inadvertently including non-HTMs in the candidates persists.
Such misclassification could lead to skewed results based on HTM events.
\begin{deluxetable*}{cccccccccccccccc}\label{tab:inject}
    \tablecaption{Injection configurations in Bayesian analysis. 
    The subscript `TP' denotes events selected from true positive candidate while `FP' denotes those from false positive candidate. The two event pairs `TP' and `FP' correspond respectively to the blue and gray $\times$ points in Fig.~\ref{fig:overlap-mock}.}
    \tablewidth{0pt}
    \tabletypesize{\footnotesize}
    \tablehead{\colhead{Parameters} & \colhead{$m_1$} & \colhead{$m_2$} & \colhead{$d_L$} & \colhead{$\chi_{1}$} & \colhead{$\chi_{2}$} & \colhead{$\theta_{1}$} & \colhead{$\theta_{2}$} & \colhead{$\phi_{\rm JL}$} & \colhead{$\phi_{1,2}$} & \colhead{$\theta$} & \colhead{$\phi$} & \colhead{$\theta_{\rm JN}$} & \colhead{$\psi$} & \colhead{$t_c$} & \colhead{$\Phi_c$}}
    \startdata
    $\rm Event_{\rm TP,1}$ & 10.4 $M_\odot$ & 9.4 $M_\odot$ & 1904 Mpc & 0.06 & 0.42 & 1.5 & 2.0 & 2.9 & 0.2 & 1.626 & 5.524 & 2.5 & 1.8 & 167021112.309 s & 0.9\\
    $\rm Event_{\rm TP,2}$ & 19.0 $M_\odot$ & 16.7 $M_\odot$ & 1904 Mpc & 0.67 & 0.39 & 2.2 & 1.6 & 4.8 & 5.8 & 1.625 & 5.524 & 2.7 & 0.5 & 198578712.309 s & 0.6\\
    $\rm Event_{\rm FP,1}$ & 24.9 $M_\odot$ & 7.7 $M_\odot$ & 11662 Mpc & 0.50 & 0.25 & 2.2 & 2.6 & 1.5 & 2.3 & 1.229 & 2.786 & 2.9 & 1.2 & 211482886.217 s & 5.5\\
    $\rm Event_{\rm FP,2}$ & 31.8 $M_\odot$ & 29.1 $M_\odot$ & 12547 Mpc & 0.44 & 0.04 & 0.5 & 1.4 & 0.4 & 1.7 & 1.245 & 2.804 & 1.4 & 1.8 & 251910294.595 s & 3.0
    \enddata
\end{deluxetable*}
\begin{figure}
    \centering
    \includegraphics[width=0.49\textwidth]{"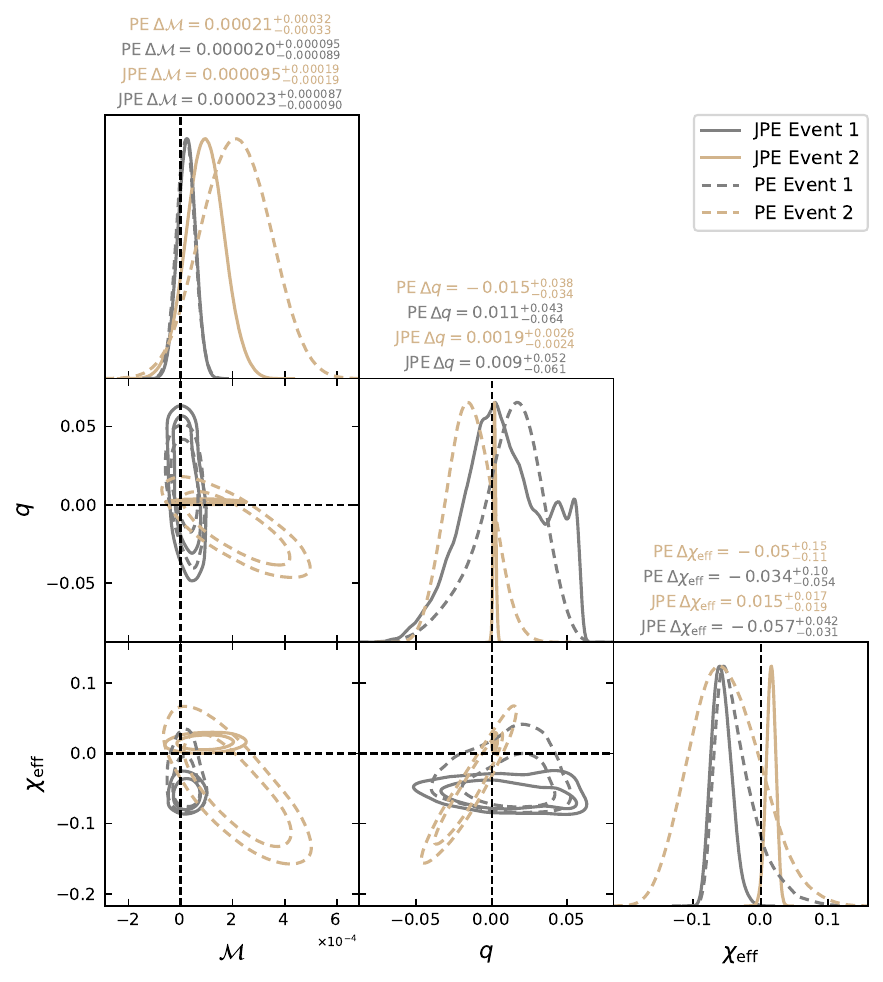"}
    \includegraphics[width=0.49\textwidth]{"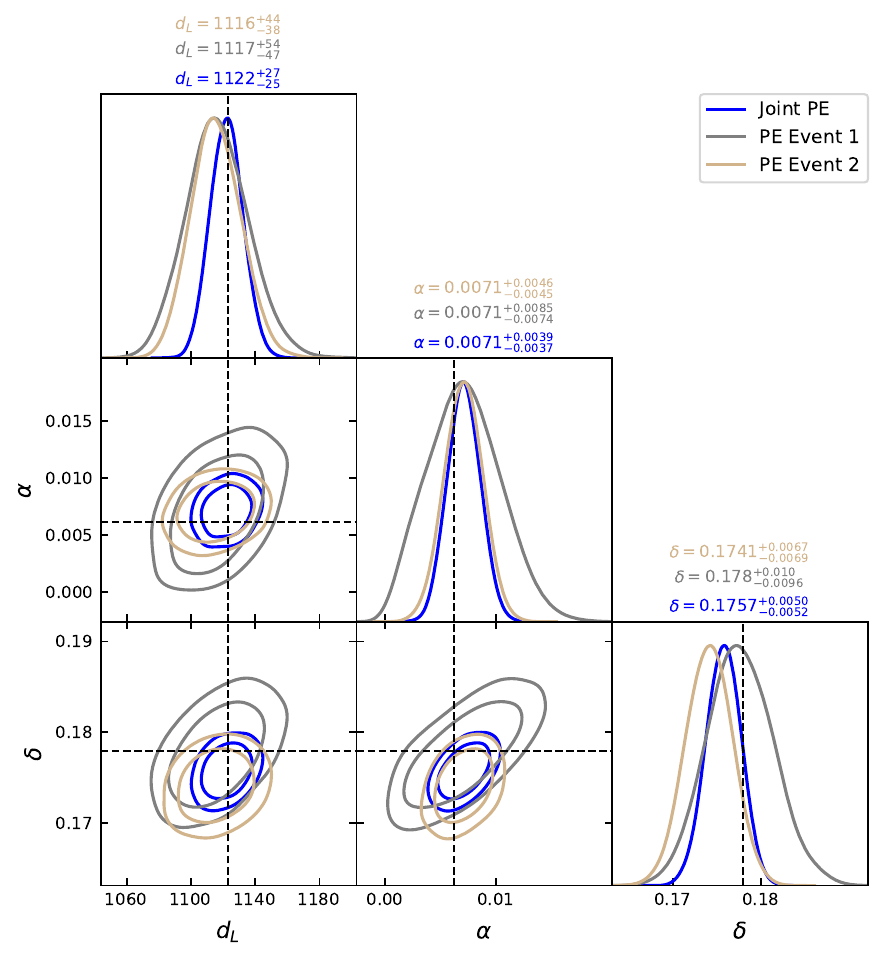"}
    \caption{
    Comparison of posterior distributions from JPE (solid lines) and individual event (dashed lines) analyses for HTM pair.
    The values reported above each diagonal plot indicate the $3\sigma$ intervals of the posterior distributions.
    The left panel presents the estimation of intrinsic parameters (the injection values have been subtracted for enhanced clarity of presentation), where the gray and tan colors represent the posterior distributions for $\rm Event_{TP,1}$ and $\rm Event_{TP,2}$, respectively.
    The right panel illustrates the estimation of extrinsic parameters, with the blue denoting the JPE analysis, while the gray and tan colors correspond to the individual event analyses for each event.}
    \label{fig:baye-comp}
\end{figure}
\begin{figure}
    \centering
    \includegraphics[width=0.49\textwidth]{"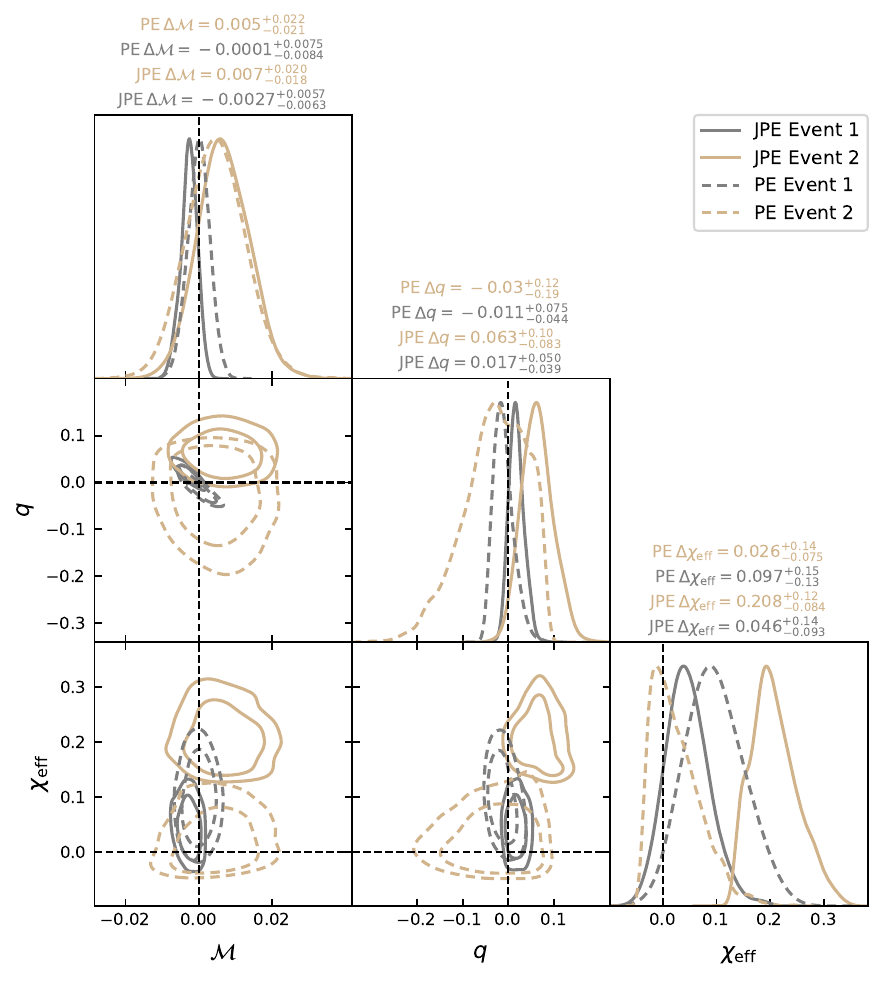"}
    \includegraphics[width=0.49\textwidth]{"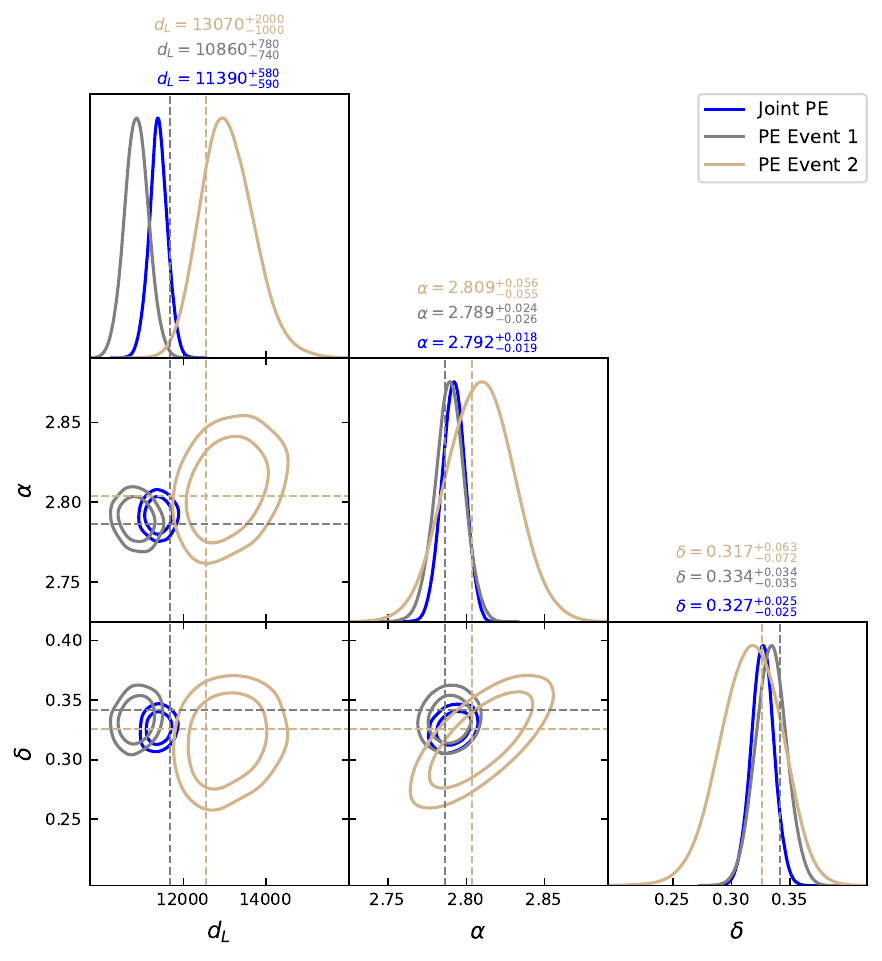"}
    \caption{Similar to Fig.~\ref{fig:baye-comp}, but is for the false positive event pair. The injection values of extrinsic parameters of $\rm Event_{\rm FP, 1}$ and $\rm Event_{\rm FP, 2}$ are represented by the gray and tan dashed lines in the right panel.}
    \label{fig:baye-com-no}
\end{figure}

Consequently, a more robust and reliable method is necessary to further validate these candidates.
To illustrate our methodology, we introduce three synthetic events to simulate the validation process, comprising one pair associated with HTM scenarios and another with non-HTM scenarios. The true values used in the injections are summarized in Table~\ref{tab:inject}. The distance from the $(1.0, 1.0)$ point in $\mathscr{M}_{\rm Bhattacharyya, Extrinsic}$ versus $\mathscr{M}_{\rm Bhattacharyya, Intrinsic}$ plane to the selected pairs of true positive and false positive events are $0.913$ and $0.909$, respectively. We conduct both JPE analyses and individual event analyses on these synthetic mergers, following the procedures outlined in Section.~\ref{subsec:JPE}.
The estimation of both extrinsic and intrinsic parameters via JPE and individual analyses is depicted in Fig.~\ref{fig:baye-comp} and Fig.~\ref{fig:baye-com-no}.
For the HTM pair, the posterior distributions from the JPE analysis align more closely with the injected parameters and exhibit reduced uncertainty compared to the results from individual analyses.
Conversely, for the non-HTM pair, JPE analysis results in poorer parameter estimation in most conditions.
We also obtain the Bayes evidence for each analysis, which allows us to calculate the Bayes factor for the two competing hypotheses using Eq.~\ref{eq:odds}.
In the case of HTM example, the logarithm of the coherence ratio is $14.5$, and the logarithm of the Odds ratio is $7.6$, thereby lending support to the HTM hypothesis.
In contrast, for the false positive non-HTM scenario, the logarithm of the coherence ratio is $7.8$, and the Odds ratio is $0.9$. Such a value lacks sufficient support for the HTM hypothesis, thereby precluding the erroneous classification of non-HTM events as HTM events.
All candidate events are subjected to the above rigorous JPE analysis to ascertain their validity. 
Ultimately, HTM events can be confidently identified through this high-credibility approach.

\section{Summary and discussion}

In the era of 3G GW detectors, we anticipate the detection of a significantly higher number of GW events, potentially exceeding 10,000 annually.
This surge in detection affords us the unprecedented opportunity to identify rare phenomena, such as HTM and lensing GW event pairs.
However, the task of distinguishing these event pairs from the multitude of detections presents a considerable challenge.
This work outlines a strategy for the identification of HTMs using 3G GW detectors.
Our approach involves a synergy between the calculation of posterior overlap and the JPE method to strike a balance between precision and computational efficiency.
Among the four methods considered for calculating the similarity between posteriors in the pre-screening phase, the Bhattacharyya coefficient method was selected for its efficacy.
The initial pre-screening of a mock dataset, containing on the order of 10,000 mergers, using the overlap calculation method can be completed in approximately one day on an Intel CPU i7-12700H.
By applying a pre-screening criterion that selects event pairs located within a unit distance from the point (1.0, 1.0), we reduced the number of potential event pairs from approximately 10 million to 23 candidates, including $70\%$ mixed HTMs.
The pre-screening process detailed in this study is demonstrably effective in both time conservation and HTM identification.
However, it is important to note that the HTM candidates may still include non-HTMs.
To address this and further validate the HTMs, JPE analyses will be conducted for each event pair among the candidates.
During this Bayesian analysis, Bayes factors are computed, serving as the decisive criterion for confirming HTMs.
Despite the time-intensive nature of Bayesian inference in the 3G era, it remains a viable approach for the analysis of 23 candidates.

Finally, we would like to explore potential improvements to this work.
Compared to the JPE method, the overlap method exhibits greater sensitivity to observational data.
The accuracy of the overlap value is influenced by the distribution of posterior and the method employed.
While the Bhattacharyya coefficient has proven effective in simulated catalogs, its performance in real 3G observation data may differ.
Comprehensive studies are necessary to ascertain the full impact of observational data on these detection strategies.
Moreover, the BBH population may affect the false alarm rate associated with the overlap method, while the JPE method might be less effective for events with low signal-to-noise ratios.
Future improvements to this research should also account for the influence of von Zeipel-Lidov-Kozai (ZLK) oscillations in hierarchical triple systems, which can introduce high eccentricities in BBH orbits.
This eccentricity could act as an indicator of triple systems and should be integrated into forthcoming identification techniques.
Additionally, some HTMs might be detectable by space-based detectors, such as LISA/Tianqin, before they enter the frequency band of terrestrial detectors, offering the prospect for multiband analyses that could enhance the study of HTMs.
The methodologies employed in the search for HTMs and strong lensing GW events are fundamentally similar.
The strategies devised for identifying HTMs could be extrapolated to the search of lensing GW signals.

\begin{acknowledgments}
The authors thank Yin-Jie Li, Chao Zhang and Zi-Jun Gao for useful discussion. This work is supported in part by the National Natural Science Foundation of China under grant Nos. U2031205, 11733009, 12233011, and 12303056, the Project for Special Research Assistant of the Chinese Academy of Sciences, and by the General Fund (No. 2023M733736) of the China Postdoctoral Science Foundation.
\end{acknowledgments}

\software{GWFAST \citep[version 1.1.1;][\url{https://github.com/CosmoStatGW/gwfast}]{2022ApJS..263....2I}, Bilby \citep[version 2.2.0][\url{https://git.ligo.org/lscsoft/bilby}]{2019ApJS..241...27A, 2020MNRAS.499.3295R}, PYCBC \citep[version 2.0.6;][\url{http://www.pycbc.org/}]{2019PASP..131b4503B}}
\bibliography{re}

\begin{thebibliography}{}
\expandafter\ifx\csname natexlab\endcsname\relax\def\natexlab#1{#1}\fi
\providecommand{\url}[1]{\href{#1}{#1}}
\providecommand{\dodoi}[1]{doi:~\href{http://doi.org/#1}{\nolinkurl{#1}}}
\providecommand{\doeprint}[1]{\href{http://ascl.net/#1}{\nolinkurl{http://ascl.net/#1}}}
\providecommand{\doarXiv}[1]{\href{https://arxiv.org/abs/#1}{\nolinkurl{https://arxiv.org/abs/#1}}}

\bibitem[{{Abbott} {et~al.}(2019){Abbott}, {Abbott}, {Abbott}, {Abraham},
  {Acernese}, {Ackley}, {Adams}, {Adhikari}, {Adya}, {Affeldt}, {Agathos},
  {Agatsuma}, {Aggarwal}, {Aguiar}, {Aiello}, {Ain}, {Ajith}, {Allen},
  {Allocca}, {Aloy}, {Altin}, {Amato}, {Ananyeva}, {Anderson}, {Anderson},
  {Angelova}, {Antier}, {Appert}, {Arai}, {Araya}, {Areeda}, {Ar{\`e}ne},
  {Arnaud}, {Arun}, {Ascenzi}, {Ashton}, {Aston}, {Astone}, {Aubin}, {Aufmuth},
  {AultONeal}, {Austin}, {Avendano}, {Avila-Alvarez}, {Babak}, {Bacon},
  {Badaracco}, {Bader}, {Bae}, {Baker}, {Baldaccini}, {Ballardin}, {Ballmer},
  {Banagiri}, {Barayoga}, {Barclay}, {Barish}, {Barker}, {Barkett}, {Barnum},
  {Barone}, {Barr}, {Barsotti}, {Barsuglia}, {Barta}, {Bartlett}, {Bartos},
  {Bassiri}, {Basti}, {Bawaj}, {Bayley}, {Bazzan}, {B{\'e}csy}, {Bejger},
  {Belahcene}, {Bell}, {Beniwal}, {Berger}, {Bergmann}, {Bernuzzi}, {Bero},
  {Berry}, {Bersanetti}, {Bertolini}, {Betzwieser}, {Bhandare}, {Bidler},
  {Bilenko}, {Bilgili}, {Billingsley}, {Birch}, {Birney}, {Birnholtz},
  {Biscans}, {Biscoveanu}, {Bisht}, {Bitossi}, {Bizouard}, {Blackburn},
  {Blair}, {Blair}, {Blair}, {Bloemen}, {Bode}, {Boer}, {Boetzel}, {Bogaert},
  {Bondu}, {Bonilla}, {Bonnand}, {Booker}, {Boom}, {Booth}, {Bork}, {Boschi},
  {Bose}, {Bossie}, {Bossilkov}, {Bosveld}, {Bouffanais}, {Bozzi},
  {Bradaschia}, {Brady}, {Bramley}, {Branchesi}, {Brau}, {Briant}, {Briggs},
  {Brighenti}, {Brillet}, {Brinkmann}, {Brisson}, {Brockill}, {Brooks},
  {Brown}, {Brunett}, {Buikema}, {Bulik}, {Bulten}, {Buonanno}, {Buscicchio},
  {Buskulic}, {Buy}, {Byer}, {Cabero}, {Cadonati}, {Cagnoli}, {Cahillane},
  {Calder{\'o}n Bustillo}, {Callister}, {Calloni}, {Camp}, {Campbell},
  {Canepa}, {Cannon}, {Cao}, {Cao}, {Capocasa}, {Carbognani}, {Caride},
  {Carney}, {Carullo}, {Casanueva Diaz}, {Casentini}, {Caudill},
  {Cavagli{\`a}}, {Cavalier}, {Cavalieri}, {Cella}, {Cerd{\'a}-Dur{\'a}n},
  {Cerretani}, {Cesarini}, {Chaibi}, {Chakravarti}, {Chamberlin}, {Chan},
  {Chao}, {Charlton}, {Chase}, {Chassande-Mottin}, {Chatterjee}, {Chaturvedi},
  {Chatziioannou}, {Cheeseboro}, {Chen}, {Chen}, {Chen}, {Cheng}, {Cheong},
  {Chia}, {Chincarini}, {Chiummo}, {Cho}, {Cho}, {Cho}, {Christensen}, {Chu},
  {Chua}, {Chung}, {Chung}, {Ciani}, {Ciobanu}, {Ciolfi}, {Cipriano}, {Cirone},
  {Clara}, {Clark}, {Clearwater}, {Cleva}, {Cocchieri}, {Coccia}, {Cohadon},
  {Cohen}, {Colgan}, {Colleoni}, {Collette}, {Collins}, {Cominsky},
  {Constancio}, {Conti}, {Cooper}, {Corban}, {Corbitt}, {Cordero-Carri{\'o}n},
  {Corley}, {Cornish}, {Corsi}, {Cortese}, {Costa}, {Cotesta}, {Coughlin},
  {Coughlin}, {Coulon}, {Countryman}, {Couvares}, {Covas}, {Cowan}, {Coward},
  {Cowart}, {Coyne}, {Coyne}, {Creighton}, {Creighton}, {Cripe}, {Croquette},
  {Crowder}, {Cullen}, {Cumming}, {Cunningham}, {Cuoco}, {Dal Canton},
  {D{\'a}lya}, {Danilishin}, {D'Antonio}, {Danzmann}, {Dasgupta}, {Da Silva
  Costa}, {Datrier}, {Dattilo}, {Dave}, {Davier}, {Davis}, {Daw}, {DeBra},
  {Deenadayalan}, {Degallaix}, {De Laurentis}, {Del{\'e}glise}, {Del Pozzo},
  {DeMarchi}, {Demos}, {Dent}, {De Pietri}, {Derby}, {De Rosa}, {De Rossi},
  {DeSalvo}, {de Varona}, {Dhurandhar}, {D{\'\i}az}, {Dietrich}, {Di Fiore},
  {Di Giovanni}, {Di Girolamo}, {Di Lieto}, {Ding}, {Di Pace}, {Di Palma}, {Di
  Renzo}, {Dmitriev}, {Doctor}, {Donovan}, {Dooley}, {Doravari}, {Dorrington},
  {Downes}, {Drago}, {Driggers}, {Du}, {Ducoin}, {Dupej}, {Dwyer}, {Easter},
  {Edo}, {Edwards}, {Effler}, {Ehrens}, {Eichholz}, {Eikenberry}, {Eisenmann},
  {Eisenstein}, {Essick}, {Estelles}, {Estevez}, {Etienne}, {Etzel}, {Evans},
  {Evans}, {Fafone}, {Fair}, {Fairhurst}, {Fan}, {Farinon}, {Farr}, {Farr},
  {Fauchon-Jones}, {Favata}, {Fays}, {Fazio}, {Fee}, {Feicht}, {Fejer}, {Feng},
  {Fernandez-Galiana}, {Ferrante}, {Ferreira}, {Ferreira}, {Ferrini},
  {Fidecaro}, {Fiori}, {Fiorucci}, {Fishbach}, {Fisher}, {Fishner},
  {Fitz-Axen}, {Flaminio}, {Fletcher}, {Flynn}, {Fong}, {Font}, {Forsyth},
  {Fournier}, {Frasca}, {Frasconi}, {Frei}, {Freise}, {Frey}, {Frey},
  {Fritschel}, {Frolov}, {Fulda}, {Fyffe}, {Gabbard}, {Gadre}, {Gaebel},
  {Gair}, {Gammaitoni}, {Ganija}, {Gaonkar}, {Garcia},
  {Garc{\'\i}a-Quir{\'o}s}, {Garufi}, {Gateley}, {Gaudio}, {Gaur}, {Gayathri},
  {Gemme}, {Genin}, {Gennai}, {George}, {George}, {Gergely}, {Germain},
  {Ghonge}, {Ghosh}, {Ghosh}, {Ghosh}, {Giacomazzo}, {Giaime}, {Giardina},
  {Giazotto}, {Gill}, {Giordano}, {Glover}, {Godwin}, {Goetz}, {Goetz},
  {Goncharov}, {Gonz{\'a}lez}, {Gonzalez Castro}, {Gopakumar}, {Gorodetsky},
  {Gossan}, {Gosselin}, {Gouaty}, {Grado}, {Graef}, {Granata}, {Grant}, {Gras},
  {Grassia}, {Gray}, {Gray}, {Greco}, {Green}, {Green}, {Gretarsson}, {Groot},
  {Grote}, {Grunewald}, {Gruning}, {Guidi}, {Gulati}, {Guo}, {Gupta}, {Gupta},
  {Gustafson}, {Gustafson}, {Haegel}, {Halim}, {Hall}, {Hall}, {Hamilton},
  {Hammond}, {Haney}, {Hanke}, {Hanks}, {Hanna}, {Hannam}, {Hannuksela},
  {Hanson}, {Hardwick}, {Haris}, {Harms}, {Harry}, {Harry}, {Haster},
  {Haughian}, {Hayes}, {Healy}, {Heidmann}, {Heintze}, {Heitmann}, {Hello},
  {Hemming}, {Hendry}, {Heng}, {Hennig}, {Heptonstall}, {Hernandez Vivanco},
  {Heurs}, {Hild}, {Hinderer}, {Hoak}, {Hochheim}, {Hofman}, {Holgado},
  {Holland}, {Holt}, {Holz}, {Hopkins}, {Horst}, {Hough}, {Howell}, {Hoy},
  {Hreibi}, {Huerta}, {Huet}, {Hughey}, {Hulko}, {Husa}, {Huttner},
  {Huynh-Dinh}, {Idzkowski}, {Iess}, {Ingram}, {Inta}, {Intini}, {Irwin},
  {Isa}, {Isac}, {Isi}, {Iyer}, {Izumi}, {Jacqmin}, {Jadhav}, {Jani},
  {Janthalur}, {Jaranowski}, {Jenkins}, {Jiang}, {Johnson}, {Jones}, {Jones},
  {Jones}, {Jonker}, {Ju}, {Junker}, {Kalaghatgi}, {Kalogera}, {Kamai},
  {Kandhasamy}, {Kang}, {Kanner}, {Kapadia}, {Karki}, {Karvinen}, {Kashyap},
  {Kasprzack}, {Katsanevas}, {Katsavounidis}, {Katzman}, {Kaufer}, {Kawabe},
  {Keerthana}, {K{\'e}f{\'e}lian}, {Keitel}, {Kennedy}, {Key}, {Khalili},
  {Khan}, {Khan}, {Khan}, {Khan}, {Khazanov}, {Khursheed}, {Kijbunchoo}, {Kim},
  {Kim}, {Kim}, {Kim}, {Kim}, {Kim}, {Kimball}, {King}, {King},
  {Kinley-Hanlon}, {Kirchhoff}, {Kissel}, {Kleybolte}, {Klika}, {Klimenko},
  {Knowles}, {Koch}, {Koehlenbeck}, {Koekoek}, {Koley}, {Kondrashov}, {Kontos},
  {Koper}, {Korobko}, {Korth}, {Kowalska}, {Kozak}, {Kringel}, {Krishnendu},
  {Kr{\'o}lak}, {Kuehn}, {Kumar}, {Kumar}, {Kumar}, {Kumar}, {Kuo}, {Kutynia},
  {Kwang}, {Lackey}, {Lai}, {Lam}, {Landry}, {Lane}, {Lang}, {Lange}, {Lantz},
  {Lanza}, {Lartaux-Vollard}, {Lasky}, {Laxen}, {Lazzarini}, {Lazzaro},
  {Leaci}, {Leavey}, {Lecoeuche}, {Lee}, {Lee}, {Lee}, {Lee}, {Lee}, {Lee},
  {Lehmann}, {Lenon}, {Leroy}, {Letendre}, {Levin}, {Li}, {Li}, {Li}, {Li},
  {Lin}, {Linde}, {Linker}, {Littenberg}, {Liu}, {Liu}, {Lo}, {Lockerbie},
  {London}, {Longo}, {Lorenzini}, {Loriette}, {Lormand}, {Losurdo}, {Lough},
  {Lousto}, {Lovelace}, {Lower}, {L{\"u}ck}, {Lumaca}, {Lundgren}, {Lynch},
  {Ma}, {Macas}, {Macfoy}, {MacInnis}, {Macleod}, {Macquet},
  {Maga{\~n}a-Sandoval}, {Maga{\~n}a Zertuche}, {Magee}, {Majorana},
  {Maksimovic}, {Malik}, {Man}, {Mandic}, {Mangano}, {Mansell}, {Manske},
  {Mantovani}, {Mapelli}, {Marchesoni}, {Marion}, {M{\'a}rka}, {M{\'a}rka},
  {Markakis}, {Markosyan}, {Markowitz}, {Maros}, {Marquina}, {Marsat},
  {Martelli}, {Martin}, {Martin}, {Martynov}, {Mason}, {Massera}, {Masserot},
  {Massinger}, {Masso-Reid}, {Mastrogiovanni}, {Matas}, {Matichard}, {Matone},
  {Mavalvala}, {Mazumder}, {McCann}, {McCarthy}, {McClelland}, {McCormick},
  {McCuller}, {McGuire}, {McIver}, {McManus}, {McRae}, {McWilliams}, {Meacher},
  {Meadors}, {Mehmet}, {Mehta}, {Meidam}, {Melatos}, {Mendell}, {Mercer},
  {Mereni}, {Merilh}, {Merzougui}, {Meshkov}, {Messenger}, {Messick},
  {Metzdorff}, {Meyers}, {Miao}, {Michel}, {Middleton}, {Mikhailov}, {Milano},
  {Miller}, {Miller}, {Millhouse}, {Mills}, {Milovich-Goff}, {Minazzoli},
  {Minenkov}, {Mishkin}, {Mishra}, {Mistry}, {Mitra}, {Mitrofanov},
  {Mitselmakher}, {Mittleman}, {Mo}, {Moffa}, {Mogushi}, {Mohapatra},
  {Montani}, {Moore}, {Moraru}, {Moreno}, {Morisaki}, {Mours}, {Mow-Lowry},
  {Mukherjee}, {Mukherjee}, {Mukherjee}, {Mukund}, {Mullavey}, {Munch},
  {Mu{\~n}iz}, {Muratore}, {Murray}, {Nagar}, {Nardecchia}, {Naticchioni},
  {Nayak}, {Neilson}, {Nelemans}, {Nelson}, {Nery}, {Neunzert}, {Ng}, {Ng},
  {Nguyen}, {Nichols}, {Nissanke}, {Nocera}, {North}, {Nuttall},
  {Obergaulinger}, {Oberling}, {O'Brien}, {O'Dea}, {Ogin}, {Oh}, {Oh}, {Ohme},
  {Ohta}, {Okada}, {Oliver}, {Oppermann}, {Oram}, {O'Reilly}, {Ormiston},
  {Ortega}, {O'Shaughnessy}, {Ossokine}, {Ottaway}, {Overmier}, {Owen}, {Pace},
  {Pagano}, {Page}, {Pai}, {Pai}, {Palamos}, {Palashov}, {Palomba},
  {Pal-Singh}, {Pan}, {Pang}, {Pang}, {Pankow}, {Pannarale}, {Pant},
  {Paoletti}, {Paoli}, {Parida}, {Parker}, {Pascucci}, {Pasqualetti},
  {Passaquieti}, {Passuello}, {Patil}, {Patricelli}, {Pearlstone}, {Pedersen},
  {Pedraza}, {Pedurand}, {Pele}, {Penn}, {Perez}, {Perreca}, {Pfeiffer},
  {Phelps}, {Phukon}, {Piccinni}, {Pichot}, {Piergiovanni}, {Pillant},
  {Pinard}, {Pirello}, {Pitkin}, {Poggiani}, {Pong}, {Ponrathnam}, {Popolizio},
  {Porter}, {Powell}, {Prajapati}, {Prasad}, {Prasai}, {Prasanna}, {Pratten},
  {Prestegard}, {Privitera}, {Prodi}, {Prokhorov}, {Puncken}, {Punturo},
  {Puppo}, {P{\"u}rrer}, {Qi}, {Quetschke}, {Quinonez}, {Quintero},
  {Quitzow-James}, {Raab}, {Radkins}, {Radulescu}, {Raffai}, {Raja}, {Rajan},
  {Rajbhandari}, {Rakhmanov}, {Ramirez}, {Ramos-Buades}, {Rana}, {Rao},
  {Rapagnani}, {Raymond}, {Razzano}, {Read}, {Regimbau}, {Rei}, {Reid},
  {Reitze}, {Ren}, {Ricci}, {Richardson}, {Richardson}, {Ricker}, {Riles},
  {Rizzo}, {Robertson}, {Robie}, {Robinet}, {Rocchi}, {Rolland}, {Rollins},
  {Roma}, {Romanelli}, {Romano}, {Romel}, {Romie}, {Rose}, {Rosi{\'n}ska},
  {Rosofsky}, {Ross}, {Rowan}, {R{\"u}diger}, {Ruggi}, {Rutins}, {Ryan},
  {Sachdev}, {Sadecki}, {Sakellariadou}, {Salconi}, {Saleem}, {Samajdar},
  {Sammut}, {Sanchez}, {Sanchez}, {Sanchis-Gual}, {Sandberg}, {Sanders},
  {Santiago}, {Sarin}, {Sassolas}, {Sathyaprakash}, {Saulson}, {Sauter},
  {Savage}, {Schale}, {Scheel}, {Scheuer}, {Schmidt}, {Schnabel}, {Schofield},
  {Sch{\"o}nbeck}, {Schreiber}, {Schulte}, {Schutz}, {Schwalbe}, {Scott},
  {Scott}, {Seidel}, {Sellers}, {Sengupta}, {Sennett}, {Sentenac}, {Sequino},
  {Sergeev}, {Setyawati}, {Shaddock}, {Shaffer}, {Shahriar}, {Shaner}, {Shao},
  {Sharma}, {Shawhan}, {Shen}, {Shink}, {Shoemaker}, {Shoemaker},
  {ShyamSundar}, {Siellez}, {Sieniawska}, {Sigg}, {Silva}, {Singer}, {Singh},
  {Singhal}, {Sintes}, {Sitmukhambetov}, {Skliris}, {Slagmolen},
  {Slaven-Blair}, {Smith}, {Smith}, {Somala}, {Son}, {Sorazu}, {Sorrentino},
  {Souradeep}, {Sowell}, {Spencer}, {Spera}, {Srivastava}, {Srivastava},
  {Staats}, {Stachie}, {Standke}, {Steer}, {Steinke}, {Steinlechner},
  {Steinlechner}, {Steinmeyer}, {Stevenson}, {Stocks}, {Stone}, {Stops},
  {Strain}, {Stratta}, {Strigin}, {Strunk}, {Sturani}, {Stuver}, {Sudhir},
  {Summerscales}, {Sun}, {Sunil}, {Suresh}, {Sutton}, {Swinkels},
  {Szczepa{\'n}czyk}, {Tacca}, {Tait}, {Talbot}, {Talukder}, {Tanner},
  {T{\'a}pai}, {Taracchini}, {Tasson}, {Taylor}, {Thies}, {Thomas}, {Thomas},
  {Thondapu}, {Thorne}, {Thrane}, {Tiwari}, {Tiwari}, {Tiwari}, {Toland},
  {Tonelli}, {Tornasi}, {Torres-Forn{\'e}}, {Torrie}, {T{\"o}yr{\"a}},
  {Travasso}, {Traylor}, {Tringali}, {Trovato}, {Trozzo}, {Trudeau}, {Tsang},
  {Tse}, {Tso}, {Tsukada}, {Tsuna}, {Tuyenbayev}, {Ueno}, {Ugolini},
  {Unnikrishnan}, {Urban}, {Usman}, {Vahlbruch}, {Vajente}, {Valdes}, {van
  Bakel}, {van Beuzekom}, {van den Brand}, {Van Den Broeck}, {Vander-Hyde},
  {van der Schaaf}, {van Heijningen}, {van Veggel}, {Vardaro}, {Varma}, {Vass},
  {Vas{\'u}th}, {Vecchio}, {Vedovato}, {Veitch}, {Veitch}, {Venkateswara},
  {Venugopalan}, {Verkindt}, {Vetrano}, {Vicer{\'e}}, {Viets}, {Vine}, {Vinet},
  {Vitale}, {Vo}, {Vocca}, {Vorvick}, {Vyatchanin}, {Wade}, {Wade}, {Wade},
  {Walet}, {Walker}, {Wallace}, {Walsh}, {Wang}, {Wang}, {Wang}, {Wang},
  {Wang}, {Ward}, {Warden}, {Warner}, {Was}, {Watchi}, {Weaver}, {Wei},
  {Weinert}, {Weinstein}, {Weiss}, {Wellmann}, {Wen}, {Wessel}, {We{\ss}els},
  {Westhouse}, {Wette}, {Whelan}, {Whiting}, {Whittle}, {Wilken}, {Williams},
  {Williamson}, {Willis}, {Willke}, {Wimmer}, {Winkler}, {Wipf}, {Wittel},
  {Woan}, {Woehler}, {Wofford}, {Worden}, {Wright}, {Wu}, {Wysocki}, {Xiao},
  {Yamamoto}, {Yancey}, {Yang}, {Yap}, {Yazback}, {Yeeles}, {Yu}, {Yu}, {Yuen},
  {Yvert}, {Zadro{\.z}ny}, {Zanolin}, {Zelenova}, {Zendri}, {Zevin}, {Zhang},
  {Zhang}, {Zhang}, {Zhao}, {Zhou}, {Zhou}, {Zhu}, {Zimmerman}, {Zlochower},
  {Zucker}, {Zweizig}, {LIGO Scientific Collaboration}, \& {Virgo
  Collaboration}}]{2019ApJ...882L..24A}
{Abbott}, B.~P., {Abbott}, R., {Abbott}, T.~D., {et~al.} 2019, \apjl, 882, L24,
  \dodoi{10.3847/2041-8213/ab3800}

\bibitem[{{Abbott} {et~al.}(2021){Abbott}, {Abbott}, {Abraham}, {Acernese},
  {Ackley}, {Adams}, {Adams}, {Adhikari}, {Adya}, {Affeldt}, \&
  et~al.}]{2021ApJ...923...14A}
{Abbott}, R., {Abbott}, T.~D., {Abraham}, S., {et~al.} 2021, \apj, 923, 14,
  \dodoi{10.3847/1538-4357/ac23db}

\bibitem[{{Abbott} {et~al.}(2023){Abbott}, {Abbott}, {Acernese}, {Ackley},
  {Adams}, {Adhikari}, {Adhikari}, {Adya}, {Affeldt}, {Agarwal}, \&
  et~al.}]{2023PhRvX..13a1048A}
{Abbott}, R., {Abbott}, T.~D., {Acernese}, F., {et~al.} 2023, Physical Review
  X, 13, 011048, \dodoi{10.1103/PhysRevX.13.011048}

\bibitem[{{Ashton} {et~al.}(2018){Ashton}, {Burns}, {Dal Canton}, {Dent},
  {Eggenstein}, {Nielsen}, {Prix}, {Was}, \& {Zhu}}]{2018ApJ...860....6A}
{Ashton}, G., {Burns}, E., {Dal Canton}, T., {et~al.} 2018, \apj, 860, 6,
  \dodoi{10.3847/1538-4357/aabfd2}

\bibitem[{{Ashton} {et~al.}(2019){Ashton}, {H{\"u}bner}, {Lasky}, {Talbot},
  {Ackley}, {Biscoveanu}, {Chu}, {Divakarla}, {Easter}, {Goncharov}, {Hernandez
  Vivanco}, {Harms}, {Lower}, {Meadors}, {Melchor}, {Payne}, {Pitkin},
  {Powell}, {Sarin}, {Smith}, \& {Thrane}}]{2019ApJS..241...27A}
{Ashton}, G., {H{\"u}bner}, M., {Lasky}, P.~D., {et~al.} 2019, \apjs, 241, 27,
  \dodoi{10.3847/1538-4365/ab06fc}

\bibitem[{Baz{\'a}n {et~al.}(2019)Baz{\'a}n, Dokl{\'a}dal, \&
  Dokl{\'a}dalov{\'a}}]{Bazn2019QuantitativeAO}
Baz{\'a}n, E., Dokl{\'a}dal, P., \& Dokl{\'a}dalov{\'a}, E. 2019, in British
  Machine Vision Conference.
\newblock \url{https://api.semanticscholar.org/CorpusID:68048254}

\bibitem[{Bhattacharyya(1946)}]{bhattacharyya1946measure}
Bhattacharyya, A. 1946, Sankhy{\=a}: the indian journal of statistics, 401

\bibitem[{{Biwer} {et~al.}(2019){Biwer}, {Capano}, {De}, {Cabero}, {Brown},
  {Nitz}, \& {Raymond}}]{2019PASP..131b4503B}
{Biwer}, C.~M., {Capano}, C.~D., {De}, S., {et~al.} 2019, \pasp, 131, 024503,
  \dodoi{10.1088/1538-3873/aaef0b}

\bibitem[{{Doctor} {et~al.}(2020){Doctor}, {Wysocki}, {O'Shaughnessy}, {Holz},
  \& {Farr}}]{2020ApJ...893...35D}
{Doctor}, Z., {Wysocki}, D., {O'Shaughnessy}, R., {Holz}, D.~E., \& {Farr}, B.
  2020, \apj, 893, 35, \dodoi{10.3847/1538-4357/ab7fac}

\bibitem[{{Evans} {et~al.}(2021){Evans}, {Adhikari}, {Afle}, {Ballmer},
  {Biscoveanu}, {Borhanian}, {Brown}, {Chen}, {Eisenstein}, {Gruson}, {Gupta},
  {Hall}, {Huxford}, {Kamai}, {Kashyap}, {Kissel}, {Kuns}, {Landry}, {Lenon},
  {Lovelace}, {McCuller}, {Ng}, {Nitz}, {Read}, {Sathyaprakash}, {Shoemaker},
  {Slagmolen}, {Smith}, {Srivastava}, {Sun}, {Vitale}, \&
  {Weiss}}]{2021arXiv210909882E}
{Evans}, M., {Adhikari}, R.~X., {Afle}, C., {et~al.} 2021, arXiv e-prints,
  arXiv:2109.09882, \dodoi{10.48550/arXiv.2109.09882}

\bibitem[{Fan {et~al.}(2023)Fan, Han, Jiang, Shao, \& Tang}]{Fan:2023spm}
Fan, Y.-Z., Han, M.-Z., Jiang, J.-L., Shao, D.-S., \& Tang, S.-P. 2023.
\newblock \doarXiv{2309.12644}

\bibitem[{{Fishbach} {et~al.}(2017){Fishbach}, {Holz}, \&
  {Farr}}]{2017ApJ...840L..24F}
{Fishbach}, M., {Holz}, D.~E., \& {Farr}, B. 2017, \apjl, 840, L24,
  \dodoi{10.3847/2041-8213/aa7045}

\bibitem[{{Flitter} {et~al.}(2021){Flitter}, {Mu{\~n}oz}, \&
  {Kovetz}}]{2021MNRAS.507..743F}
{Flitter}, J., {Mu{\~n}oz}, J.~B., \& {Kovetz}, E.~D. 2021, \mnras, 507, 743,
  \dodoi{10.1093/mnras/stab2203}

\bibitem[{{Gerosa} \& {Berti}(2017)}]{2017PhRvD..95l4046G}
{Gerosa}, D., \& {Berti}, E. 2017, \prd, 95, 124046,
  \dodoi{10.1103/PhysRevD.95.124046}

\bibitem[{{Gerosa} \& {Fishbach}(2021)}]{2021NatAs...5..749G}
{Gerosa}, D., \& {Fishbach}, M. 2021, Nature Astronomy, 5, 749,
  \dodoi{10.1038/s41550-021-01398-w}

\bibitem[{{Haris} {et~al.}(2018){Haris}, {Mehta}, {Kumar}, {Venumadhav}, \&
  {Ajith}}]{2018arXiv180707062H}
{Haris}, K., {Mehta}, A.~K., {Kumar}, S., {Venumadhav}, T., \& {Ajith}, P.
  2018, arXiv e-prints, arXiv:1807.07062, \dodoi{10.48550/arXiv.1807.07062}

\bibitem[{{Hild} {et~al.}(2011){Hild}, {Abernathy}, {Acernese}, {Amaro-Seoane},
  {Andersson}, {Arun}, {Barone}, {Barr}, {Barsuglia}, {Beker}, {Beveridge},
  {Birindelli}, {Bose}, {Bosi}, {Braccini}, {Bradaschia}, {Bulik}, {Calloni},
  {Cella}, {Chassande Mottin}, {Chelkowski}, {Chincarini}, {Clark}, {Coccia},
  {Colacino}, {Colas}, {Cumming}, {Cunningham}, {Cuoco}, {Danilishin},
  {Danzmann}, {De Salvo}, {Dent}, {De Rosa}, {Di Fiore}, {Di Virgilio},
  {Doets}, {Fafone}, {Falferi}, {Flaminio}, {Franc}, {Frasconi}, {Freise},
  {Friedrich}, {Fulda}, {Gair}, {Gemme}, {Genin}, {Gennai}, {Giazotto},
  {Glampedakis}, {Gr{\"a}f}, {Granata}, {Grote}, {Guidi}, {Gurkovsky},
  {Hammond}, {Hannam}, {Harms}, {Heinert}, {Hendry}, {Heng}, {Hennes}, {Hough},
  {Husa}, {Huttner}, {Jones}, {Khalili}, {Kokeyama}, {Kokkotas}, {Krishnan},
  {Li}, {Lorenzini}, {L{\"u}ck}, {Majorana}, {Mandel}, {Mandic}, {Mantovani},
  {Martin}, {Michel}, {Minenkov}, {Morgado}, {Mosca}, {Mours},
  {M{\"u}ller{\textendash}Ebhardt}, {Murray}, {Nawrodt}, {Nelson},
  {Oshaughnessy}, {Ott}, {Palomba}, {Paoli}, {Parguez}, {Pasqualetti},
  {Passaquieti}, {Passuello}, {Pinard}, {Plastino}, {Poggiani}, {Popolizio},
  {Prato}, {Punturo}, {Puppo}, {Rabeling}, {Rapagnani}, {Read}, {Regimbau},
  {Rehbein}, {Reid}, {Ricci}, {Richard}, {Rocchi}, {Rowan}, {R{\"u}diger},
  {Santamar{\'\i}a}, {Sassolas}, {Sathyaprakash}, {Schnabel}, {Schwarz},
  {Seidel}, {Sintes}, {Somiya}, {Speirits}, {Strain}, {Strigin}, {Sutton},
  {Tarabrin}, {Th{\"u}ring}, {van den Brand}, {van Veggel}, {van den Broeck},
  {Vecchio}, {Veitch}, {Vetrano}, {Vicere}, {Vyatchanin}, {Willke}, {Woan}, \&
  {Yamamoto}}]{2011CQGra..28i4013H}
{Hild}, S., {Abernathy}, M., {Acernese}, F., {et~al.} 2011, Classical and
  Quantum Gravity, 28, 094013, \dodoi{10.1088/0264-9381/28/9/094013}

\bibitem[{Iacovelli {et~al.}(2022{\natexlab{a}})Iacovelli, Mancarella, Foffa,
  \& Maggiore}]{Iacovelli:2022mbg}
Iacovelli, F., Mancarella, M., Foffa, S., \& Maggiore, M. 2022{\natexlab{a}},
  Astrophys. J. Supp., 263, 2, \dodoi{10.3847/1538-4365/ac9129}

\bibitem[{Iacovelli {et~al.}(2022{\natexlab{b}})Iacovelli, Mancarella, Foffa,
  \& Maggiore}]{Iacovelli:2022bbs}
---. 2022{\natexlab{b}}, Astrophys. J., 941, 208,
  \dodoi{10.3847/1538-4357/ac9cd4}

\bibitem[{{Iacovelli} {et~al.}(2022){Iacovelli}, {Mancarella}, {Foffa}, \&
  {Maggiore}}]{2022ApJS..263....2I}
{Iacovelli}, F., {Mancarella}, M., {Foffa}, S., \& {Maggiore}, M. 2022, \apjs,
  263, 2, \dodoi{10.3847/1538-4365/ac9129}

\bibitem[{{Janquart} {et~al.}(2023{\natexlab{a}}){Janquart}, {Baka},
  {Samajdar}, {Dietrich}, \& {Van Den Broeck}}]{2023MNRAS.523.1699J}
{Janquart}, J., {Baka}, T., {Samajdar}, A., {Dietrich}, T., \& {Van Den
  Broeck}, C. 2023{\natexlab{a}}, \mnras, 523, 1699,
  \dodoi{10.1093/mnras/stad1542}

\bibitem[{{Janquart} {et~al.}(2023{\natexlab{b}}){Janquart}, {Haris},
  {Hannuksela}, \& {Van Den Broeck}}]{2023MNRAS.526.3088J}
{Janquart}, J., {Haris}, K., {Hannuksela}, O.~A., \& {Van Den Broeck}, C.
  2023{\natexlab{b}}, \mnras, 526, 3088, \dodoi{10.1093/mnras/stad2838}

\bibitem[{{Janquart} {et~al.}(2023{\natexlab{c}}){Janquart}, {More}, \& {Van
  Den Broeck}}]{2023MNRAS.519.2046J}
{Janquart}, J., {More}, A., \& {Van Den Broeck}, C. 2023{\natexlab{c}}, \mnras,
  519, 2046, \dodoi{10.1093/mnras/stac3660}

\bibitem[{{Krishna} {et~al.}(2023){Krishna}, {Vijaykumar}, {Ganguly}, {Talbot},
  {Biscoveanu}, {George}, {Williams}, \& {Zimmerman}}]{2023arXiv231206009K}
{Krishna}, K., {Vijaykumar}, A., {Ganguly}, A., {et~al.} 2023, arXiv e-prints,
  arXiv:2312.06009.
\newblock \doarXiv{2312.06009}

\bibitem[{{L{\'a}zaro-Gredilla} {et~al.}(2012){L{\'a}zaro-Gredilla}, {Van
  Vaerenbergh}, \& {Lawrence}}]{2012PatRe..45.1386L}
{L{\'a}zaro-Gredilla}, M., {Van Vaerenbergh}, S., \& {Lawrence}, N.~D. 2012,
  Pattern Recognition, 45, 1386, \dodoi{10.1016/j.patcog.2011.10.004}

\bibitem[{{Li} {et~al.}(2023){Li}, {Wang}, {Tang}, \&
  {Fan}}]{2023arXiv230302973L}
{Li}, Y.-J., {Wang}, Y.-Z., {Tang}, S.-P., \& {Fan}, Y.-Z. 2023, arXiv
  e-prints, arXiv:2303.02973, \dodoi{10.48550/arXiv.2303.02973}

\bibitem[{Liang {et~al.}(2017)Liang, Wang, Wang, Li, Hu, Jin, Fan, Liang, \&
  Wei}]{Liang:2017cjo}
Liang, Y.-F., Wang, Y.-Z., Wang, H., {et~al.} 2017.
\newblock \doarXiv{1705.01881}

\bibitem[{{Liu} {et~al.}(2023){Liu}, {Wang}, {Hu}, {Tanikawa}, \&
  {Trani}}]{2023arXiv231105393L}
{Liu}, S., {Wang}, L., {Hu}, Y.-M., {Tanikawa}, A., \& {Trani}, A.~A. 2023,
  arXiv e-prints, arXiv:2311.05393, \dodoi{10.48550/arXiv.2311.05393}

\bibitem[{{Liu} {et~al.}(2021){Liu}, {Maga{\~n}a Hernandez}, \&
  {Creighton}}]{2021ApJ...908...97L}
{Liu}, X., {Maga{\~n}a Hernandez}, I., \& {Creighton}, J. 2021, \apj, 908, 97,
  \dodoi{10.3847/1538-4357/abd7eb}

\bibitem[{{Mandel} \& {Farmer}(2022)}]{2022PhR...955....1M}
{Mandel}, I., \& {Farmer}, A. 2022, \physrep, 955, 1,
  \dodoi{10.1016/j.physrep.2022.01.003}

\bibitem[{{Mapelli} {et~al.}(2021){Mapelli}, {Santoliquido}, {Bouffanais},
  {Arca Sedda}, {Artale}, \& {Ballone}}]{2021Symm...13.1678M}
{Mapelli}, M., {Santoliquido}, F., {Bouffanais}, Y., {et~al.} 2021, Symmetry,
  13, 1678, \dodoi{10.3390/sym13091678}

\bibitem[{{Miller} \& {Hamilton}(2002)}]{2002MNRAS.330..232C}
{Miller}, M.~C., \& {Hamilton}, D.~P. 2002, \mnras, 330, 232,
  \dodoi{10.1046/j.1365-8711.2002.05112.x}

\bibitem[{{Oancea} {et~al.}(2023){Oancea}, {Stiskalek}, \&
  {Zumalac{\'a}rregui}}]{2023arXiv230701903O}
{Oancea}, M.~A., {Stiskalek}, R., \& {Zumalac{\'a}rregui}, M. 2023, arXiv
  e-prints, arXiv:2307.01903, \dodoi{10.48550/arXiv.2307.01903}

\bibitem[{Ouali {et~al.}(2020)Ouali, Mahdi, Gharbaoui, \& Medjahed}]{article2D}
Ouali, M., Mahdi, W., Gharbaoui, R., \& Medjahed, S.~A. 2020, Computación y
  Sistemas, 24, \dodoi{10.13053/cys-24-3-3326}

\bibitem[{{Planck Collaboration} {et~al.}(2020){Planck Collaboration},
  {Aghanim}, {Akrami}, {Ashdown}, {Aumont}, {Baccigalupi}, {Ballardini},
  {Banday}, {Barreiro}, {Bartolo}, {Basak}, {Battye}, {Benabed}, {Bernard},
  {Bersanelli}, {Bielewicz}, {Bock}, {Bond}, {Borrill}, {Bouchet}, {Boulanger},
  {Bucher}, {Burigana}, {Butler}, {Calabrese}, {Cardoso}, {Carron},
  {Challinor}, {Chiang}, {Chluba}, {Colombo}, {Combet}, {Contreras}, {Crill},
  {Cuttaia}, {de Bernardis}, {de Zotti}, {Delabrouille}, {Delouis}, {Di
  Valentino}, {Diego}, {Dor{\'e}}, {Douspis}, {Ducout}, {Dupac}, {Dusini},
  {Efstathiou}, {Elsner}, {En{\ss}lin}, {Eriksen}, {Fantaye}, {Farhang},
  {Fergusson}, {Fernandez-Cobos}, {Finelli}, {Forastieri}, {Frailis},
  {Fraisse}, {Franceschi}, {Frolov}, {Galeotta}, {Galli}, {Ganga},
  {G{\'e}nova-Santos}, {Gerbino}, {Ghosh}, {Gonz{\'a}lez-Nuevo}, {G{\'o}rski},
  {Gratton}, {Gruppuso}, {Gudmundsson}, {Hamann}, {Handley}, {Hansen},
  {Herranz}, {Hildebrandt}, {Hivon}, {Huang}, {Jaffe}, {Jones}, {Karakci},
  {Keih{\"a}nen}, {Keskitalo}, {Kiiveri}, {Kim}, {Kisner}, {Knox},
  {Krachmalnicoff}, {Kunz}, {Kurki-Suonio}, {Lagache}, {Lamarre}, {Lasenby},
  {Lattanzi}, {Lawrence}, {Le Jeune}, {Lemos}, {Lesgourgues}, {Levrier},
  {Lewis}, {Liguori}, {Lilje}, {Lilley}, {Lindholm}, {L{\'o}pez-Caniego},
  {Lubin}, {Ma}, {Mac{\'\i}as-P{\'e}rez}, {Maggio}, {Maino}, {Mandolesi},
  {Mangilli}, {Marcos-Caballero}, {Maris}, {Martin}, {Martinelli},
  {Mart{\'\i}nez-Gonz{\'a}lez}, {Matarrese}, {Mauri}, {McEwen}, {Meinhold},
  {Melchiorri}, {Mennella}, {Migliaccio}, {Millea}, {Mitra},
  {Miville-Desch{\^e}nes}, {Molinari}, {Montier}, {Morgante}, {Moss}, {Natoli},
  {N{\o}rgaard-Nielsen}, {Pagano}, {Paoletti}, {Partridge}, {Patanchon},
  {Peiris}, {Perrotta}, {Pettorino}, {Piacentini}, {Polastri}, {Polenta},
  {Puget}, {Rachen}, {Reinecke}, {Remazeilles}, {Renzi}, {Rocha}, {Rosset},
  {Roudier}, {Rubi{\~n}o-Mart{\'\i}n}, {Ruiz-Granados}, {Salvati}, {Sandri},
  {Savelainen}, {Scott}, {Shellard}, {Sirignano}, {Sirri}, {Spencer},
  {Sunyaev}, {Suur-Uski}, {Tauber}, {Tavagnacco}, {Tenti}, {Toffolatti},
  {Tomasi}, {Trombetti}, {Valenziano}, {Valiviita}, {Van Tent}, {Vibert},
  {Vielva}, {Villa}, {Vittorio}, {Wandelt}, {Wehus}, {White}, {White},
  {Zacchei}, \& {Zonca}}]{2020A&A...641A...6P}
{Planck Collaboration}, {Aghanim}, N., {Akrami}, Y., {et~al.} 2020, \aap, 641,
  A6, \dodoi{10.1051/0004-6361/201833910}

\bibitem[{{Pratten} {et~al.}(2021){Pratten}, {Garc{\'\i}a-Quir{\'o}s},
  {Colleoni}, {Ramos-Buades}, {Estell{\'e}s}, {Mateu-Lucena}, {Jaume}, {Haney},
  {Keitel}, {Thompson}, \& {Husa}}]{2021PhRvD.103j4056P}
{Pratten}, G., {Garc{\'\i}a-Quir{\'o}s}, C., {Colleoni}, M., {et~al.} 2021,
  \prd, 103, 104056, \dodoi{10.1103/PhysRevD.103.104056}

\bibitem[{{Punturo} {et~al.}(2010){Punturo}, {Abernathy}, {Acernese}, {Allen},
  {Andersson}, {Arun}, {Barone}, {Barr}, {Barsuglia}, {Beker}, {Beveridge},
  {Birindelli}, {Bose}, {Bosi}, {Braccini}, {Bradaschia}, {Bulik}, {Calloni},
  {Cella}, {Chassande Mottin}, {Chelkowski}, {Chincarini}, {Clark}, {Coccia},
  {Colacino}, {Colas}, {Cumming}, {Cunningham}, {Cuoco}, {Danilishin},
  {Danzmann}, {De Luca}, {De Salvo}, {Dent}, {De Rosa}, {Di Fiore}, {Di
  Virgilio}, {Doets}, {Fafone}, {Falferi}, {Flaminio}, {Franc}, {Frasconi},
  {Freise}, {Fulda}, {Gair}, {Gemme}, {Gennai}, {Giazotto}, {Glampedakis},
  {Granata}, {Grote}, {Guidi}, {Hammond}, {Hannam}, {Harms}, {Heinert},
  {Hendry}, {Heng}, {Hennes}, {Hild}, {Hough}, {Husa}, {Huttner}, {Jones},
  {Khalili}, {Kokeyama}, {Kokkotas}, {Krishnan}, {Lorenzini}, {L{\"u}ck},
  {Majorana}, {Mandel}, {Mandic}, {Martin}, {Michel}, {Minenkov}, {Morgado},
  {Mosca}, {Mours}, {M{\"u}ller{\textendash}Ebhardt}, {Murray}, {Nawrodt},
  {Nelson}, {Oshaughnessy}, {Ott}, {Palomba}, {Paoli}, {Parguez},
  {Pasqualetti}, {Passaquieti}, {Passuello}, {Pinard}, {Poggiani}, {Popolizio},
  {Prato}, {Puppo}, {Rabeling}, {Rapagnani}, {Read}, {Regimbau}, {Rehbein},
  {Reid}, {Rezzolla}, {Ricci}, {Richard}, {Rocchi}, {Rowan}, {R{\"u}diger},
  {Sassolas}, {Sathyaprakash}, {Schnabel}, {Schwarz}, {Seidel}, {Sintes},
  {Somiya}, {Speirits}, {Strain}, {Strigin}, {Sutton}, {Tarabrin},
  {Th{\"u}ring}, {van den Brand}, {van Leewen}, {van Veggel}, {van den Broeck},
  {Vecchio}, {Veitch}, {Vetrano}, {Vicere}, {Vyatchanin}, {Willke}, {Woan},
  {Wolfango}, \& {Yamamoto}}]{2010CQGra..27s4002P}
{Punturo}, M., {Abernathy}, M., {Acernese}, F., {et~al.} 2010, Classical and
  Quantum Gravity, 27, 194002, \dodoi{10.1088/0264-9381/27/19/194002}

\bibitem[{{Reitze} {et~al.}(2019){Reitze}, {Adhikari}, {Ballmer}, {Barish},
  {Barsotti}, {Billingsley}, {Brown}, {Chen}, {Coyne}, {Eisenstein}, {Evans},
  {Fritschel}, {Hall}, {Lazzarini}, {Lovelace}, {Read}, {Sathyaprakash},
  {Shoemaker}, {Smith}, {Torrie}, {Vitale}, {Weiss}, {Wipf}, \&
  {Zucker}}]{2019BAAS...51g..35R}
{Reitze}, D., {Adhikari}, R.~X., {Ballmer}, S., {et~al.} 2019, in Bulletin of
  the American Astronomical Society, Vol.~51, 35,
  \dodoi{10.48550/arXiv.1907.04833}

\bibitem[{{Romero-Shaw} {et~al.}(2020){Romero-Shaw}, {Talbot}, {Biscoveanu},
  {D'Emilio}, {Ashton}, {Berry}, {Coughlin}, {Galaudage}, {Hoy}, {H{\"u}bner},
  {Phukon}, {Pitkin}, {Rizzo}, {Sarin}, {Smith}, {Stevenson}, {Vajpeyi},
  {Ar{\`e}ne}, {Athar}, {Banagiri}, {Bose}, {Carney}, {Chatziioannou}, {Clark},
  {Colleoni}, {Cotesta}, {Edelman}, {Estell{\'e}s}, {Garc{\'\i}a-Quir{\'o}s},
  {Ghosh}, {Green}, {Haster}, {Husa}, {Keitel}, {Kim}, {Hernandez-Vivanco},
  {Maga{\~n}a Hernandez}, {Karathanasis}, {Lasky}, {De Lillo}, {Lower},
  {Macleod}, {Mateu-Lucena}, {Miller}, {Millhouse}, {Morisaki}, {Oh},
  {Ossokine}, {Payne}, {Powell}, {Pratten}, {P{\"u}rrer}, {Ramos-Buades},
  {Raymond}, {Thrane}, {Veitch}, {Williams}, {Williams}, \&
  {Xiao}}]{2020MNRAS.499.3295R}
{Romero-Shaw}, I.~M., {Talbot}, C., {Biscoveanu}, S., {et~al.} 2020, \mnras,
  499, 3295, \dodoi{10.1093/mnras/staa2850}

\bibitem[{{Samsing} \& {Ilan}(2019)}]{2019MNRAS.482...30S}
{Samsing}, J., \& {Ilan}, T. 2019, \mnras, 482, 30,
  \dodoi{10.1093/mnras/sty2249}

\bibitem[{{Samsing} {et~al.}(2022){Samsing}, {Bartos}, {D'Orazio}, {Haiman},
  {Kocsis}, {Leigh}, {Liu}, {Pessah}, \& {Tagawa}}]{2022Natur.603..237S}
{Samsing}, J., {Bartos}, I., {D'Orazio}, D.~J., {et~al.} 2022, \nat, 603, 237,
  \dodoi{10.1038/s41586-021-04333-1}

\bibitem[{{Tang} {et~al.}(2023){Tang}, {Fan}, \& {Wei}}]{2023MNRAS.523.4113T}
{Tang}, S.-P., {Fan}, Y.-Z., \& {Wei}, D.-M. 2023, \mnras, 523, 4113,
  \dodoi{10.1093/mnras/stad1676}

\bibitem[{Tang {et~al.}(2024)Tang, Gao, Li, Fan, \& Wei}]{Tang:2023zxa}
Tang, S.-P., Gao, B., Li, Y.-J., Fan, Y.-Z., \& Wei, D.-M. 2024, Astrophys. J.,
  960, 67, \dodoi{10.3847/1538-4357/ad0dfa}

\bibitem[{{The LIGO Scientific Collaboration} {et~al.}(2021{\natexlab{a}}){The
  LIGO Scientific Collaboration}, {the Virgo Collaboration}, {the KAGRA
  Collaboration}, {Abbott}, {Abe}, {Acernese}, {Ackley}, {Adhikari},
  {Adhikari}, {Adkins}, \& et~al.}]{2021arXiv211206861T}
{The LIGO Scientific Collaboration}, {the Virgo Collaboration}, {the KAGRA
  Collaboration}, {et~al.} 2021{\natexlab{a}}, arXiv e-prints,
  arXiv:2112.06861, \dodoi{10.48550/arXiv.2112.06861}

\bibitem[{{The LIGO Scientific Collaboration} {et~al.}(2021{\natexlab{b}}){The
  LIGO Scientific Collaboration}, {the Virgo Collaboration}, {the KAGRA
  Collaboration}, {Abbott}, {Abbott}, {Acernese}, {Ackley}, {Adams},
  {Adhikari}, {Adhikari}, \& et~al.}]{2021arXiv211103606T}
---. 2021{\natexlab{b}}, arXiv e-prints, arXiv:2111.03606,
  \dodoi{10.48550/arXiv.2111.03606}

\bibitem[{{The LIGO Scientific Collaboration} {et~al.}(2023){The LIGO
  Scientific Collaboration}, {the Virgo Collaboration}, {the KAGRA
  Collaboration}, {Abbott}, {Abe}, {Acernese}, {Ackley}, {Adhicary},
  {Adhikari}, {Adhikari}, \& et~al.}]{2023arXiv230408393T}
---. 2023, arXiv e-prints, arXiv:2304.08393, \dodoi{10.48550/arXiv.2304.08393}

\bibitem[{{Trani} {et~al.}(2022){Trani}, {Rastello}, {Di Carlo},
  {Santoliquido}, {Tanikawa}, \& {Mapelli}}]{2022MNRAS.511.1362T}
{Trani}, A.~A., {Rastello}, S., {Di Carlo}, U.~N., {et~al.} 2022, \mnras, 511,
  1362, \dodoi{10.1093/mnras/stac122}

\bibitem[{{Veske} {et~al.}(2020){Veske}, {M{\'a}rka}, {Sullivan}, {Bartos},
  {Rainer Corley}, {Samsing}, \& {M{\'a}rka}}]{2020MNRAS.498L..46V}
{Veske}, D., {M{\'a}rka}, Z., {Sullivan}, A.~G., {et~al.} 2020, \mnras, 498,
  L46, \dodoi{10.1093/mnrasl/slaa123}

\bibitem[{{Veske} {et~al.}(2021){Veske}, {Sullivan}, {M{\'a}rka}, {Bartos},
  {Corley}, {Samsing}, {Buscicchio}, \& {M{\'a}rka}}]{2021ApJ...907L..48V}
{Veske}, D., {Sullivan}, A.~G., {M{\'a}rka}, Z., {et~al.} 2021, \apjl, 907,
  L48, \dodoi{10.3847/2041-8213/abd721}

\bibitem[{{Zackay} {et~al.}(2018){Zackay}, {Dai}, \&
  {Venumadhav}}]{2018arXiv180608792Z}
{Zackay}, B., {Dai}, L., \& {Venumadhav}, T. 2018, arXiv e-prints,
  arXiv:1806.08792, \dodoi{10.48550/arXiv.1806.08792}

\end{thebibliography}
\bibliographystyle{aasjournal}
\end{document}